\documentclass[11pt]{article}
\usepackage{fullpage,enumitem,afterpage,setspace,graphicx,amsfonts,array,verbatim,commath,mathrsfs}
\usepackage{authblk}

\RequirePackage{amsthm,amsmath,amsfonts,amssymb}
\RequirePackage[authoryear]{natbib}
\RequirePackage[colorlinks,citecolor=blue,urlcolor=blue]{hyperref}
\RequirePackage{graphicx}
\usepackage{mathtools}
\usepackage{bm}
\usepackage{bbm}
\usepackage{booktabs}
\usepackage{makecell}
\usepackage{algorithm}
\usepackage{algorithmic}
\usepackage{siunitx}
\usepackage{caption}
\usepackage{sectsty}
\usepackage[title]{appendix}
\usepackage{url}

\numberwithin{equation}{section}

\DeclareMathOperator*{\argmax}{argmax}

\newcommand{\changed}[1]{\textcolor{black}{#1}}

\def\Var{\text{Var}}
\def\E{{\mathbb E}}
\def\med{\text{median}}

\begin{document}

\singlespacing

\title{A flexible model for correlated count data, \\ with application to multi-condition differential expression analyses of single-cell RNA sequencing data}
\author[1]{Yusha Liu}
\author[2,3]{Peter Carbonetto}
\author[4,5]{Michihiro Takahama}
\author[6]{Adam Gruenbaum}
\author[7]{Dongyue Xie}
\author[4]{Nicolas Chevrier}
\author[2,7]{Matthew Stephens\thanks{{\tt mstephens@uchicago.edu}}}
\affil[1]{\small Department of Biostatistics, The University of North Carolina at Chapel Hill, Chapel Hill, NC, USA}
\affil[2]{\small Department of Human Genetics, The University of Chicago, Chicago, IL, USA}
\affil[3]{\small Research Computing Center, The University of Chicago, Chicago, IL, USA}
\affil[4]{\small Pritzker School of Molecular Engineering, The University of Chicago, Chicago, IL, USA}
\affil[5]{\small Graduate School of Pharmaceutical Sciences, Osaka University, Osaka, Japan}
\affil[6]{\small Institute for Health Metrics and Evaluation, University of Washington, Seattle, WA, USA} 
\affil[7]{\small Department of Statistics, The University of Chicago, Chicago, IL, USA}
\date{}
\maketitle

\begin{abstract}
Detecting differences in gene expression is an important part of
single-cell RNA sequencing experiments, and many statistical methods
have been developed for this aim. Most differential expression
analyses focus on comparing expression between two groups (e.g.,
treatment vs. control). But there is increasing interest in {\em
  multi-condition differential expression analyses} in which
expression is measured in many conditions and the aim is to
accurately detect and estimate expression differences in all
conditions. We show that directly modeling single-cell RNA-seq counts
in all conditions simultaneously, while also inferring how expression
differences are shared across conditions, leads to greatly improved
performance for detecting and estimating expression differences
compared to existing methods.
We illustrate the potential of this new approach by analyzing data
from a single-cell experiment studying the effects of cytokine
stimulation on gene expression. We call our new method ``Poisson
multivariate adaptive shrinkage'', and it is implemented in an R
package available at
\url{https://github.com/stephenslab/poisson.mash.alpha}.
\end{abstract}

\section{Introduction}
\label{sec:intro}

Detecting differences in gene expression --- that is, ``differential
expression'' (DE) analysis --- has been a fundamental analysis aim
ever since the introduction of technologies to measure gene expression
\citep{soneson2013comparison}.
As measurement technologies have improved, gene expression data sets
have increased in size and resolution, bringing new analysis
challenges. The development of RNA sequencing (RNA-seq) technologies
\citep{wang2009rnaseq}
has greatly facilitated the measurement of gene expression in ``bulk''
samples.  More recently, the development of single-cell RNA sequencing
technologies (scRNA-seq) has allowed for rapid, high-throughput
measurement of gene expression in individual cells, resulting in large
datasets profiling gene expression in thousands of cells
\citep[e.g.,][]{zheng2017massively}.

Increasingly, biologists are developing scRNA-seq experiments in which
gene expression is assayed {\em in many experimental conditions.}  For
example, in the motivating data set for this paper, gene expression
data were obtained for approximately 142,000 cells
under 45 different treatment conditions. In {\em multi-condition data}
such as these, we seek to understand which changes in expression are
specific to certain conditions (``condition-specific effects''), and
which changes are shared among two or more conditions (``shared
effects'').  In this paper, we develop methods to tackle these aims
--- specifically, to detect which genes are differentially expressed,
and to estimate the log-fold changes (LFCs) {\em among multiple
  conditions.} While many methods exist for performing differential
expression analysis of scRNA-seq data, analyzing multi-condition
scRNA-seq data raises at least two key challenges that are not
adequately addressed by existing methods.

First, when assessing expression across multiple conditions, many
different patterns of differential expression are possible. For
example, some genes may be differentially expressed in a single
condition (relative to all other conditions), while other genes may
show similar expression differences in subsets of conditions.
Typically, these patterns are unknown in advance, but one would like
to identify and exploit them to improve accuracy of the LFC estimates,
and to improve power to detect differentially expressed genes. To
address this first challenge, we build on the empirical Bayes (EB)
method developed in \cite{urbut2019flexible}, ``multivariate adaptive
shrinkage'' (``mash'' for short), which is designed to model and
adapt to effect-sharing patterns among conditions present in the data.

Second, the data from scRNA-seq experiments are molecular counts,
which are most naturally modeled using count models such as Poisson
measurement models \citep{townes2019feature, sarkar2021separating}.
However, there is no straightforward way to integrate a Poisson model
with mash because the Poisson model does not naturally provide summary
statistics --- effect estimates and standard errors --- that can be
used by mash; in particular, estimates of standard errors are
unreliable in Poisson models \citep{robinson2008small}. An alternative
would be to combine mash with a Gaussian measurement model for {\em
  log-transformed} scRNA-seq counts \citep[e.g.,][]{finak2015mast}.
However, as has been repeatedly pointed out \citep[e.g.,][]{lun2018overcoming,
townes2019feature, muscat},
this data transformation can lead to severe bias in the LFC
estimates. This is particularly an issue when many of the counts are
zero or small, {\em which is a common feature of scRNA-seq data sets.}
This suggests that it would be desirable to combine mash with a model
of the scRNA-seq counts.

Therefore, to get the best of both worlds --- improved accuracy
achieved by exploiting patterns of effect-sharing across conditions
and the advantages of directly modeling the scRNA-seq counts without
first transforming them --- we pursue an approach that {\em models the
  scRNA-seq count data jointly in all conditions.}  We call this new
approach ``Poisson mash'' because it is based on a Poisson model of
the data, and, like mash, it improves accuracy in the effect estimates
by flexibly modeling the sharing of effects across conditions.  Since
the gains in accuracy will be greater as more conditions with shared
effects are included in the analysis, in this paper we focus on
scRNA-seq experiments in which gene expression is measured in many
(e.g., dozens) conditions. 
Although its development has been motivated by our interest in
analyzing multi-condition scRNA-seq data sets, the Poisson mash model
can also be viewed as a general model of multivariate count data, so
the ideas contained in this paper may be useful in other settings
where multivariate count data occur.

The structure of the paper is as follows. First, in Section
\ref{sec:problem}, we define ``multi-condition differential expression
analysis'' more formally, and explain the underlying assumptions about
the data. Next, we introduce the core Poisson mash model (Section
\ref{sec:basic}), discuss related methods (Section \ref{sec:related}),
then describe several enhancements to the model that improve its
performance in more realistic settings (Section \ref{sec:extensions}).
In Section \ref{sec:simulations}, we evaluate the benefits of Poisson
mash approach compared with existing methods in simulated scRNA-seq
data sets. To illustrate how Poisson mash can be used to gain
biological insights from multi-condition gene expression data, we
apply Poisson mash to the scRNA-seq data set mentioned above (Section
\ref{sec:application}). Finally, we wrap up with a discussion (Section
\ref{sec:discussion}).

\subsection{Software availability}

The Poisson mash methods are implemented in the R package 
 ``poisson.mash.alpha'', which is available at
\url{https://github.com/stephenslab/poisson.mash.alpha}.

\section{Problem setup}
\label{sec:problem}

In a {\em multi-condition differential expression analysis},
  the aim is to compare expression for each of $J$ genes across 
  $R$ conditions from multi-condition count data $\bm{X}$,
\begin{equation}
{\bm X} =\; \rotatebox[origin=c]{90}{\mbox{$J$ genes}} 
\underset{\mbox{$R$ conditions}}{\left[\begin{array}{cccc}
x_{11} & x_{12} & \cdots & x_{1R} \\
x_{21} & x_{22} & \cdots & x_{2R} \\
\vdots & \vdots & \ddots & \vdots \\
x_{J1} & x_{J2} & \cdots & x_{JR}
\end{array}\right]}.
\end{equation}
This matrix can be obtained by summing, for each gene $j$, the unique
molecular identifier (UMI) counts from all cells in the same condition
(see Section \ref{sec:pseudobulk}). The special case of $R = 2$ ---
that is, when ${\bm X}$ is a $J \times 2$ matrix --- corresponds to
the standard setup for DE analysis in which the aim is to compare
expression between two conditions (e.g., treatment vs. control). By
contrast, we focus on {\em multi-condition experiments with $R \gg
  2$}; for example, in the cytokines data, $R = 45$.

Next, we assume a {\em Poisson measurement model} for the counts,
\begin{equation}
x_{jr} \sim \mathrm{Pois}(s_r \lambda_{jr}), 
\label{eq:pois}
\end{equation}
independently for each gene $j$ and condition $r$, in which $s_r > 0$
denotes a ``size factor''.\footnote{In DE analyses of scRNA-seq
  data, the size factors $s_r$ are often defined as the sum of the
  counts in each condition, $s_r = \sum_{j=1}^J x_{jr}$. This is the
  default setting for our analyses, noting that other
  definitions are possible \citep[e.g.,][]{bullard2010evaluation}.}
The parameter $\lambda_{jr}$ in \eqref{eq:pois} represents a relative
rate, specifically the {\em relative expression level} for gene $j$ in
condition $r$.

To analyze differences in expression, the log-relative expression is
decomposed into a baseline level of expression, $\mu_j$, and a
condition-specific expression difference $\beta_{jr}$ relative to the
baseline:
\begin{equation}
\log \lambda_{jr} = \mu_{j} + \beta_{jr}. 
\label{eq:lambda}
\end{equation}
Our main aim is to accurately estimate the expression differences
$\beta_{jr}$ and other statistical quantities involving
$\beta_{jr}$. With these modeling assumptions, the $\mu_j$ and
$\beta_{jr}$ are not individually identifiable from the count data,
${\bm X}$. However, they become identifiable once one introduces
priors for $\beta_{jr}$; this is described in the next section where
we introduce the Poisson mash model.

The assumption that the data are in the form of a $J \times R$
matrix ${\bm X}$ implies that there is only a {\em single independent
  observation per condition.}
This assumption is not critical; when multiple replicates are
available per condition, we can aggregate cells by replicate rather
than by condition go that there are multiple independent
observations per condition. This extension is described in
Appendix \ref{sec:replicate}. But the
single-observation assumption simplifies the description of the
method, and furthermore the extension to multiple observations is not
essential for understanding of the method nor appreciating its
benefits.

\subsection[Obtaining X from multi-condition scRNA-seq data]{\changed{Obtaining ${\bm X}$ from multi-condition scRNA-seq data}} 
\label{sec:pseudobulk}

In multi-condition scRNA-seq data, we observe the UMI counts $y_{ji}$
for genes $j \in \{1, \ldots, J\}$ in cells $i \in \{1, \ldots, N\}$
in which each cell $i$ is measured in one of $R$ conditions. To
analyze the differences in expression across $R$ conditions following
the setup described above, we summarize the UMI counts in each
condition by summing them across cells from the same condition; that
is, for each gene $j$ and condition $r$, we set $x_{jr} = \sum_{i
  \,\in\, \mathcal{S}_r} y_{ji}$, where $\mathcal{S}_r \subset \{1,
\ldots, N\}$ denotes the indices of the cells that are measured in
condition $r$.

The idea of summing the UMI counts from the individual cells is
commonly described as ``pseudobulk analysis'', and its benefits were
noted in several recent papers \citep{lun2017overcoming,
  ahlmann2020glmgampoi, erdmann-pham2021likelihoodbased,
  murphy2022balanced, muscat, squair2021confronting}. In the context
of multi-condition DE analysis, forming pseudobulk data has the twin
advantages of simplifying modeling and reducing computation, and for
these reasons we take this approach here. One possible concern with a
pseudobulk analysis of single-cell RNA-seq data is that one may need
to correct for unwanted variation (known or unknown) that must be
accounted for at the single-cell level \citep{ruvseq,
  leek2007capturing, svaseq, gerard2020empirical}.
We address this concern in Section \ref{sec:ruv}.

\section{The basic Poisson mash model} 
\label{sec:basic}

We now give the minimum details needed to understand Poisson mash.
Enhancements to the basic model are described in Section
\ref{sec:extensions}.

\subsection{The multivariate adaptive shrinkage prior}

Our main aim is to detect and estimate expression differences among
conditions. For example, if some subset of conditions involves
treatments with similar biological effects, then the expression
differences $\beta_{jr}$ are expected to be similar to one another; on
the other hand, if one treatment has a very different biological
effect from other treatments, it may show a ``condition-specific''
effect in which $\beta_{jr}$ is nonzero only in that
condition. Furthermore, the patterns of DE may vary across genes; for
example, genes in the same pathway may show more similar patterns of
DE than genes in different pathways. In summary, different data sets
will likely exhibit different patterns of DE among conditions, and
multiple patterns of DE may be present within a single data set.

\begin{figure}
\centering
\includegraphics[width=0.75\textwidth]{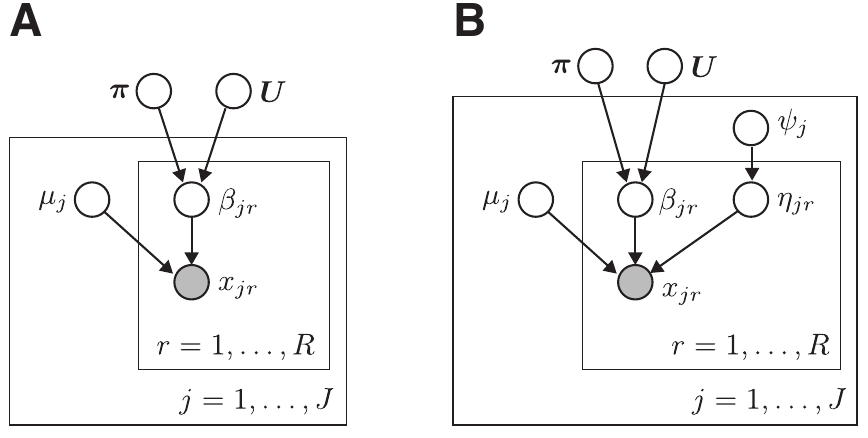}
\caption{\rm Directed acyclic graphs (DAGs) showing the
    independence structure of the basic Poisson mash model (A) and the
    Poisson mash model augmented with random effects (B).}
\label{fig:dags}
\end{figure}

To capture heterogeneous DE patterns, and adapt these patterns to the
data, we use the multivariate adaptive shrinkage (``mash'') prior
introduced in \cite{urbut2019flexible},
\begin{equation}
p(\bm{\beta}_j; {\bm\pi}, {\bm U}) = 
\sum_{k=1}^K \sum_{l=1}^L \pi_{kl} N_R(\bm{\beta}_j;
\bm{0}, w_l {\bm U}_k), \label{eq:mash_prior}
\end{equation}
where ${\bm\beta}_j \coloneqq (\beta_{j1}, \ldots, \beta_{jR})'$, and
$N_R(\,\cdot\,; \bm{\mu}, \bm\Sigma)$ denotes the density of the
$R$-variate normal distribution with mean $\bm{\mu}$ and covariance
matrix $\bm\Sigma$. Here, $w_l > 0, l = 1, \dots, L$, is a
pre-specified ``grid'' of scaling coefficients, spanning from very
small to very large, to capture the full range of possible effect
sizes, and the $\pi_{kl}$ are mixture weights, $\pi_{kl} \geq 0$,
$\sum_{k=1}^K \sum_{l=1}^L \pi_{kl} = 1$. Each ${\bm U}_k$ is an $R
\times R$ covariance matrix that captures a pattern of covariation of
effects across conditions. We refer to the Poisson measurement model
(\ref{eq:pois}--\ref{eq:lambda}) together with the mash prior
\eqref{eq:mash_prior} as ``Poisson mash''. The graphical model
representation of the Poisson mash model is shown in Panel A of
Figure~\ref{fig:dags}.

The covariance matrices ${\bm U}_k$ can include both pre-specified
``canonical'' covariance matrices that represent, for example, DE
specific to a condition, and ``data-driven'' covariance matrices that
are estimated from the data, and can capture arbitrary patterns of DE
among the conditions. Each weight $\pi_{kl}$ should capture the
relative frequency of each combination of scaling coefficient $w_l$
and cross-condition pattern ${\bm U}_k$. These weights will be
estimated from the data.

The mash prior \eqref{eq:mash_prior} is centered on zero, which helps
address the nonidentifiability of $\mu_j$ and $\beta_{jr}$ in
\eqref{eq:lambda}; specifically, centering the prior on zero
encourages the average $\beta_{jr}$ to be near zero, and is analogous
to the (non-Bayesian) approach of identifying parameters by imposing the
constraint $\sum_{r=1}^R \beta_{jr}=0$. In addition, we note that
typically the prior will place considerable weight near zero,
reflecting the fact that many expression differences are zero or
small. This has the effect of ``shrinking'' many of the estimated
$\beta_{jr}$ towards zero.

Our model is closely connected to multivariate Poisson log-normal
(MPLN) distributions, originally proposed by
\cite{aitchison1989multivariate} as a flexible approach to modeling
dependencies among Poisson variables.  Indeed, integrating out the 
effects $\bm{\beta}_j$, the marginal distribution of the counts for
gene $j$, ${\bm x}_j \coloneqq (x_{j1}, \dots, x_{jR})'$, is a mixture
of MPLNs. Similar MPLN mixture models have recently been used to
cluster multivariate count measurements in biological applications
\citep{silva2019multivariate, subedi2020family}.

\subsection{Inference aims}
\label{sec:inference}

With the Poisson mash model, we focus on the following types of
inferences:

\paragraph*{Inference Aim 1: Detect differentially expressed genes} 

Test the ``global null'' hypothesis of no DE for gene $j$, $H_{0j}$:
$\beta_{jr} = 0 \; \forall r$.

\paragraph*{Inference Aim 2: Detect and estimate expression 
  differences relative to a reference condition}

When a single condition serves as the control condition, we detect and
estimate changes in each condition $r$ relative to the control condition,
and quantify uncertainty in these estimates, say, by reporting
interval estimates. For example, supposing condition 1 is the control,
then $\log(\lambda_{jr}/\lambda_{j1}) = \beta_{jr} - \beta_{j1}$ is
the log-fold change in condition $r \geq 2$ relative to the
control.\footnote{The convention in DE analysis, originating from DE
analyses in microarray experiments, is to use the base-2 logarithm.
In all our results, we report LFC estimates and related quantities
using the base-2 logarithm.}
  
Not all experiments have a single natural reference or control
condition. In such cases, one could estimate changes in condition $r$
relative to the mean or median across all conditions: $\beta_{jr} -
\mathrm{mean}\{\beta_{j1}, \ldots, \beta_{jR}\}$ or $\beta_{jr} -
\mathrm{median}\{\beta_{j1}, \ldots, \beta_{jR}\}$.  The median may be
preferable to the mean in situations when a few conditions are very
different from the others. For example, suppose the expression of gene
$j$ is upregulated in condition $r$ only. In this example, the
deviation of $\beta_{jr}$ from the median will be nonzero (and
positive) only in condition $r$, whereas its deviation from the mean
will be positive in condition $r$ and negative (and smaller in
magnitude) in all other conditions.
  
\subsection{Variational empirical Bayes algorithm}

We take the same empirical Bayes (EB) approach in mash
\citep{urbut2019flexible}. This involves two key computations:
\renewcommand{\labelenumi}{(\roman{enumi})} \setlength{\itemsep}{6pt}
\begin{enumerate}

\item Estimate the mash prior --- specifically, the prior covariances
  ${\bm U} \coloneq \{{\bm U}_1, \ldots, {\bm U}_K\}$ and the mixture
  weights ${\bm \pi}$ --- by pooling information from the $J$ genes.

\item Compute gene-specific posterior quantities using the 
  prior estimated in (i).

\end{enumerate} 

A major difference between mash and Poisson mash is that mash is based
on a Gaussian measurement model, and therefore benefits from the
convenient analytical properties of mixtures of multivariate
Gaussians, whereas Poisson mash is based on a Poisson measurement
model that does not result in analytic posterior computations.
Therefore, some approximations must be made to obtain computations
that are analytic and tractable. We propose to use variational
approximation techniques \citep{blei2017variational}
to obtain fast posterior computations.  Specifically, we 
use the Gaussian variational approximation described by
\cite{arridge2018variational}. With this approximation, the exact
posterior distribution of $\bm{\beta}_j$,
\begin{align} 
\label{eq:post_exact}
p_{\mathrm{post}}({\bm\beta}_j) \coloneqq&\;
p(\bm{\beta}_j \mid {\bm x}_j, \mu_j, \bm{\pi}, {\bm U}) \nonumber \\
\propto&\; p({\bm x}_j \mid \bm{\beta}_j, \mu_j) \, 
p(\bm{\beta}_j; {\bm\pi}, {\bm U}) \nonumber \\
\propto&\; p({\bm x}_j \mid \bm{\beta}_j, \mu_j) 
\times \sum_{k=1}^K \sum_{l=1}^L  \pi_{kl} \: N_R(\bm{\beta}_j;
  \bm{0}, w_l {\bm U}_k),
\end{align}
which does not have an analytic formula, is approximated by a mixture
of multivariate Gaussian distributions,
\begin{equation} 
\label{eq:post_approx}
p_{\mathrm{post}}({\bm\beta}_j) \approx q_j({\bm\beta}_j) 
\coloneqq \sum_{k=1}^K \sum_{l=1}^L \zeta_{jkl} 
N_R({\bm\beta}_j; {\bm\varphi}_{jkl},  {\bm\Sigma}_{jkl}).
\end{equation}
That is, $q_j$ acts as the approximate posterior for
$\bm{\beta}_j$. The idea behind the variational inference approach is
to search for the free parameters $\zeta_{jkl}, {\bm\varphi}_{jkl},
{\bm\Sigma}_{jkl}$, $k = 1, \ldots, K$, $l = 1, \ldots, L$, that
produce a $q_j$ which most closely resembles the true posterior
distribution of ${\bm\beta}_j$. This parameter search is most
frequently done using numerical optimization techniques.  That is,
when we describe fitting the approximate posterior distribution $q_j$,
we are actually optimizing the parameters $\zeta_{jkl},
{\bm\varphi}_{jkl}, {\bm\Sigma}_{jkl}$.

To develop algorithms for mixture models, such as EM algorithms, a
common data augmentation trick is to introduce a latent variable that
indicates the source mixture component. For each gene $j$, we define a
latent indicator $z_{jkl} \in \{0, 1\}$ that is 1 if ${\bm\beta}_j$ is
drawn from mixture component $(j, k)$, and zero otherwise. With this
data augmentation, we have
\begin{equation}
\bm{\beta}_j \mid z_{jkl} = 1 \sim N_R(\bm{0}, w_l \bm{U}_k),
\label{eq:z}
\end{equation} 
and the approximate posterior is defined as
\begin{equation}
\label{eq:post_approx_augmented}
q_j(\bm{\beta}_j, \bm{z}_j) = 
\prod_{k=1}^K \prod_{l=1}^L \big\{ \zeta_{jkl} 
N_R(\bm{\beta}_j; \bm{\varphi}_{jkl},  \bm{\Sigma}_{jkl}) 
\big\}^{z_{jkl}}.
\end{equation}
This definition recovers \eqref{eq:post_approx} after marginalizing
over ${\bm z}_j$.

The algorithm for fitting the variational approximation
proceeds by maximizing a lower bound to the likelihood, sometimes
called the ``evidence lower bound'', or ELBO
\citep{blei2017variational}.
Since the likelihood naturally factorizes over the genes, the ELBO in
turn is a simple sum,
\begin{equation} 
\label{eq:elbo}
\mathrm{ELBO}(q_1, \ldots, q_J; {\bm\mu}, \bm{\pi}, {\bm U}) =
\sum_{j=1}^J \mathrm{ELBO}_j(q_j; \mu_j, {\bm\pi}, {\bm U}),
\end{equation}
in which the ELBO for gene $j$ is
\begin{equation}
\label{eq:geneelbo}
\mathrm{ELBO}_j(q_j; \mu_j, {\bm\pi}, {\bm U}) \coloneqq 
\log p(\bm{x}_j \mid \mu_j, \bm{\pi}, \bm{U}) - 
D_{\mathrm{KL}}(q_j(\bm{\beta}_j, {\bm z}_j) \,\|\, 
p_{\mathrm{post}}(\bm{\beta}_j, {\bm z}_j)),
\end{equation}
where $D_{\mathrm{KL}}(q \,\|\, p)$ denotes the Kullback-Leibler (KL)
divergence from $q$ to $p$ \citep{cover-thomas}. While both terms on
the right-hand side of \eqref{eq:geneelbo} are intractable, the
intractable parts will cancel out in the expression for the ELBO,
which leads to tractable computations under the approximation
\eqref{eq:post_approx}. See Appendix \ref{sec:fitting} for detailed derivations.

To maximize the ELBO \eqref{eq:elbo}, we take an ``EM-like''
co-ordinate ascent approach, in which we alternate between maximizing
the ELBO with respect to the approximate posteriors $q_1, \ldots, q_J$
(the ``E step''), and maximizing the ELBO with respect to the model
parameters ${\bm\mu}, \bm{\pi}, \bm{U}$ (the ``M step''), until some
convergence criterion is met, yielding parameter estimates
$\hat{\bm\mu}, \hat{\bm{\pi}}, \hat{\bm{U}}$ and approximate posteriors
$\hat{q}_1, \ldots, \hat{q}_J$. This co-ordinate ascent
algorithm resembles an EM algorithm except that the E step produces
approximate posterior expectations, and for this reason this algorithm
is sometimes called ``variational EM''
\citep{blei2017variational}. Note that the M step in this algorithm
pools the information across all $J$ genes to estimate the parameters,
whereas the posterior computations in the E step are independent for
each gene $j$ and therefore can be performed in parallel. For more
details on this, see Appendix \ref{sec:fitting}.

\subsection{Posterior statistics}

Now we define the posterior quantities that we use to tackle the
inference aims, and we briefly explain how they are computed.

We estimate expression differences and quantify their uncertainty
(Inference Aim 2) using the approximate posteriors
$\hat{q}_j(\bm{\beta}_j)$. Since the approximate posteriors are
mixtures of multivariate normals, posterior means and covariances of
$\bm{\beta}_{j}$ are available analytically, and other posterior
quantities can be computed by Monte Carlo simulation. In particular,
we estimate the LFC relative to a control (assumed without loss of
generality to be condition 1) as simply the posterior expectation
$\E[\beta_{jr} - \beta_{j1}] = \E[\beta_{jr}] - \E[\beta_{j1}]$, in
which these expectations are taken with respect to the approximate
posterior $\hat{q}_j$. When the LFC is defined relative to the median
expression level, we compute a Monte Carlo estimate by simulating from
$\hat{q}_j$.

To determine whether an expression difference is significant
(Inference Aim 2), we use the {\em local false sign rate} ({\em lfsr})
\citep{stephens2017false},
\begin{equation}  
\label{eq:lfsr}
\mbox{\em lfsr}_{jr} \coloneqq 
\min \left\{ \Pr(\beta_{jr} \geq 0), \Pr(\beta_{jr} \leq 0) \right\},
\end{equation}
in which the probabilities $\Pr(\beta_{jr} \geq 0)$ and
$\Pr(\beta_{jr} \leq 0)$ are again obtained from $\hat{q}_j$.  The {\em
  lfsr} is a measure of significance that is analogous to local false
discovery rate ({\em lfdr}) but more conservative since it controls
the probability that the sign of $\beta_{jr}$ is incorrectly estimated
rather than the probability that $\beta_{jr}$ is incorrectly called
nonzero (given the observed data). The {\em lfsr} was also found to be
more robust to modeling assumptions than the {\em lfdr}
\citep{stephens2017false}.

 The {\em lfsr} \eqref{eq:lfsr} defines a {\em condition-specific}
 measure of significance; to tackle Inference Aim 1, we
   need to define a {\em gene-level} measure of significance. To this
   end, we propose the {\em minimum lfsr},
\begin{equation}
\label{eq:minlfsr}
\mbox{\em min-lfsr}_j \coloneqq \min\{ \mbox{\em lfsr}_{j1}, \ldots, 
\mbox{\em lfsr}_{jR}\}.
\end{equation}
Gene $j$ is considered to be a differentially expressed gene
  if $\mbox{\em min-lfsr}_j < \alpha$, and is considered to be
  differentially expressed in condition $r$ if $\mbox{\em lfsr}_{jr} <
  \alpha$, for some $\alpha \in (0,1)$. The {\em lfsr} threshold
  $\alpha$ controls the stringency of the tests and is chosen by the
  analyst; in this paper, we use $\alpha = 0.05$ unless stated
  otherwise.

We note that a simpler alternative to the {\em minimum lfsr} is to
compute a {\em Bayes factor}
for ${\bm\beta}_j \ne {\bm 0}$ vs. ${\bm\beta}_j = {\bm 0}$,
\begin{equation*}
\mathrm{BF}_j \coloneqq
\frac{p({\bm x}_j \mid \hat\mu_j, \bm{\hat\pi}, \bm{\hat{U}})}
     {p({\bm x}_j \mid \hat{\mu}_j^{\mathrm{null}}, \bm{\beta}_j = \bm{0)}},
\end{equation*}
where $\hat{\mu}_j^{\mathrm{null}}$ is an estimate of $\mu_j$ under
the null model. In practice, the numerator in the Bayes factor is
difficult to compute and therefore we approximate it by the ELBO,
$\log p({\bm x}_j \mid \hat\mu_j, \bm{\hat\pi}, \bm{\hat{U}}) \approx
\mathrm{ELBO}_j(\hat{q}_j; \hat{\mu}_j, \hat{\bm\pi}, \hat{\bm
  U})$. The ELBO is a lower bound, so the approximate Bayes factor
will always be an underestimate of the exact Bayes factor, resulting
in more conservative detection of DE genes. In our initial
evaluations, we found that the Bayes factor did not perform as well as
the {\em minimum lfsr}, perhaps due to the approximation used in our
calculation of the Bayes factors, so we recommend using the {\em
  minimum lfsr} for a gene-level measure of significance.

We also implemented a test to assess the goodness-of-fit for a Poisson
mash model; the details are given in Appendix \ref{sec:goodness-of-fit}.

\section{Related work} 
\label{sec:related}

Many statistical methods are available to perform multi-condition DE
analysis of scRNA-seq data. Extensive reviews and comparisons of these
methods have been conducted \citep{soneson2018bias,
  wang2019comparative} and we refer the reader to these papers for
details on the available methods. Among the variety of DE analysis
methods, widely used methods include limma \citep{law2014voom,
  smyth2004linear}, MAST \citep{finak2015mast}, edgeR
\citep{robinson2010edger} and DESeq2 \citep{deseq, love2014moderated,
  zhu2019heavy}. While limma, edgeR and DESeq2 were originally
developed respectively for microarray expression data and bulk RNA-seq
data, they have also been found to work well for scRNA-seq data
\citep{soneson2018bias}. MAST, by contrast, was specifically designed
to cope with the particulars of scRNA-seq data. None of these methods
share the ability of Poisson mash to combine information across
conditions.
  
MultiDE \citep{kang2016multide} and CorMotif \citep{wei2015joint},
which were developed with bulk RNA-seq data in mind, share some of the
features of Poisson mash. Among the two, MultiDE is more like Poisson
mash. Like Poisson mash, MultiDE is aimed at performing DE analysis
jointly over multiple conditions. MultiDE models RNA-seq counts
$y_{ji}$ using the negative binomial (NB) distribution; for $i \in
\mathcal{S}_r, y_{ji} \sim \mbox{NB}(\mu_{jr}, \phi_j)$, in which $j$
indexes genes, $r$ indexes conditions and $i$ indexes replicates. A
key difference is that MultiDE makes a restrictive assumption about
how expression differences are shared across conditions. Specifically,
MultiDE assumes $\log \mu_{jr} = \mu_j + u_r v_j$. Therefore, MultiDE
can be viewed as a special case of the Poisson mash model in which the
mash prior \eqref{eq:mash_prior} has a single component ($K = 1$), and
a single covariance matrix, ${\bm U}_1=\bm u \bm u'$, where $\bm
u=(u_1,\dots,u_R)'$. MultiDE is not expected to perform well when
these prior assumptions are violated.

CorMotif is intended for comparing differences in expression between
two groups in multiple independent studies (or conditions, or
cell-types, etc.). A similar setup is also considered in a recent
benchmarking paper assessing methods for differential expression
analysis \citep{muscat}.  However, this setup is quite different from
our setup where we have a single observation in each of the many
conditions, and our aim is to compare gene expression across
conditions. Additionally, in contrast with Poisson mash, CorMotif
focuses on DE detection, and not on estimation of the effect sizes
(log-fold changes), and CorMotif does not directly model the counts.
Despite these differences, CorMotif shares with Poisson mash the idea
of modeling shared patterns of DE across multiple studies (they call
these patterns ``correlation motifs''), and using these shared
patterns to increase power of DE detection.

\section{Enhancements to the basic Poisson mash model} 
\label{sec:extensions}

In this section, we describe two important practical improvements to
the basic Poisson mash model: (i) a ``random effect'' to account for
additional sources of experimental variation; and (ii) latent factors
to account for ``unwanted variation'' that can induce dependence among
the gene-level tests. (In the remainder of this paper, the random
effect enhancement is included in all applications of the method
unless stated otherwise.)

\subsection{Modeling random effects} 
\label{sec:random_effects}

Even in the absence of expression differences ($\beta_{jr} = 0$ for
all $r = 1, \ldots, R$), there might be heterogeneity in
$\lambda_{jr}$ across conditions due to other sources of
experimental variation. To account for this variation, we introduce a
``random effect'' $\eta_{jr}$,
\begin{equation}
\log \lambda_{jr} = \mu_{j} + \beta_{jr} + \eta_{jr},
\label{eq:random_effect}
\end{equation}
and assign a normal prior to the random effects for each gene,
\begin{equation}
\bm{\eta}_j \sim N_R(\bm{0}, \psi_j^2 {\bm I}_R), 
\label{eq:random_effect_prior}
\end{equation}
where ${\bm\eta}_j \coloneqq (\eta_{j1}, \ldots, \eta_{jR})'$,
$\psi_j^2$ is an unknown, gene-specific parameter to be estimated from
the data, and ${\bm I}_R$ is the $R \times R$ identity matrix. The
basic model \eqref{eq:lambda} is recovered from this model by setting
$\psi_j^2 = 0$. Panel B of Figure~\ref{fig:dags} shows the 
graphical model for Poisson mash with random effects.

Based solely on the Poisson likelihood, only the sum $\beta_{jr} +
\eta_{jr}$ is identifiable, not the individual terms in this sum. The
different priors are what make it possible to simultaneously estimate
both the expression differences $\beta_{jr}$ and the random effects
$\eta_{jr}$;
the prior for $\bm{\eta}_{j}$ is independent across conditions,
whereas the prior for $\bm{\beta}_j$ is not. Therefore, correlations
across conditions are explained only by the expression differences
$\beta_{jr}$ and not by the random effects $\eta_{jr}$. In addition,
because $\eta_{jr}$ has an identical normal prior in all conditions,
the model may prefer to explain a strong condition-specific effect in
condition $r$ using $\beta_{jr}$ rather than $\eta_{jr}$.

The addition of the random effect in \eqref{eq:random_effect} can be
seen as an alternative to the negative binomial model (as used by
edgeR and DESeq2, as well as other methods for analyzing RNA-seq data)
to allow for greater flexibility in modeling variation in the
counts. In particular, by integrating out $\bm{\eta}_j$, $x_{jr}$ is
marginally modeled by a {\em Poisson-log normal distribution}. The
mean and variance of the counts under this model are
\begin{align*}
\E[x_{jr}] &= s_r e^{\mu_j + \beta_{jr} + \psi_j^2/2}  \\
\Var[x_{jr}] &= 
\E[x_{jr}] \times \big\{1 + \E[x_{jr}](e^{\psi_j^2} - 1)\big\}. 
\end{align*}
It is easy to see from these expressions that $\psi_j^2$ affects the
level of overdispersion for gene $j$, and in particular, when
$\psi_j^2 = 0$ there is no overdispersion; that is, $\E[x_{jr}] =
\Var[x_{jr}]$, recovering the property of the Poisson that its mean
and variance are the same. 

We note that the Poisson log-normal model was used in
\cite{gu2014badge} to model inter-sample variation for DE analysis of
RNA-seq data.

\subsection{Correcting for unwanted variation} 
\label{sec:ruv}

We further extend the Poisson mash model to allow for the
incorporation of $D$ additional variables ${\bm \rho}_d \coloneqq
(\rho_{1d}, \ldots, \rho_{Rd})'$, $d = 1, \ldots, D$. These variables
represent sources of unwanted variation present in the data that can
induce dependence among gene-wise tests and confound DE analysis
\citep{leek2007capturing, svaseq}.
The augmented model is
\begin{equation}
\log \lambda_{jr} = 
\mu_{j} + \beta_{jr} + \eta_{jr} + \sum_{d=1}^D f_{jd} \rho_{rd},
\label{eqn:general_pm} 
\end{equation}
in which the $f_{jd}$ are the regression coefficients for the
confounding variables. Let ${\bm F}$ denote the $J \times D$ matrix
with elements $f_{jd}$, and let ${\bm\rho}$ denote the $R \times D$
matrix with elements $\rho_{rd}$. We assume ${\bm F}$ has been
previously estimated from the cell-level data ${\bm Y}$, and is
therefore treated as ``known'' when fitting the Poisson mash model,
whereas ${\bm\rho}$ will be estimated along with the other parameters
of the Poisson mash model.

Because the primary motivation for including these additional
variables is to correct for unwanted variation, we call this augmented
model ``Poisson mash RUV'', where the ``RUV'' is short for ``removing
unwanted variation''. In our experiments (Section
\ref{sec:simulations}), we compare Poisson mash with and without the
RUV enhancement --- that is, the models based on the two different
definitions of the Poisson rates $\lambda_{jr}$ in
\eqref{eq:random_effect} and in \eqref{eqn:general_pm} --- to show how
correcting for unwanted variation improves detection of expression
differences.

In some cases, the confounding variables are known, such as when these
capture batch effects. More often, the unwanted variation is due to
unmeasured factors, in which case these variables can be estimated
using one of the several methods developed for this aim, such as
RUVSeq \citep{ruvseq}, svaseq \citep{leek2007capturing, svaseq} or
mouthwash \citep{gerard2020empirical}.
(The best method to use may depend on the
specifics of the RNA-seq experiment, e.g., whether negative controls
are available.) Regardless of the specific method used, ${\bm X}$
alone cannot be used to estimate the unknown confounders.  When the
cell-level data ${\bm Y}$ are available, it is reasonable to estimate
${\bm F}$ from ${\bm Y}$,
then use the same ${\bm F}$ for Poisson mash (for justification, see
Appendix \ref{sec:ruv_appendix}).

In our analyses, we estimated ${\bm F}$ by fitting a GLM-PCA model
\citep{townes2019feature} with $D$ factors to the single-cell data
(the UMI counts) ${\bm Y}$. To avoid learning factors that correspond
to the conditions, we included the condition labels as covariates in
the GLM-PCA model. After fitting the GLM-PCA model, we took ${\bm F}$
to be the ${\bm V}$ matrix from the GLM-PCA model fit (following the
notation of \citealt{townes2019feature}).

\section{Simulations} 
\label{sec:simulations}

We performed simulations to evaluate Poisson mash and compare with
other DE analysis approaches.

Most comparisons of DE analysis methods have focused on evaluating
their ability to detect differentially expressed genes (e.g.,
\citealt{squair2021confronting, soneson2018bias,
  wang2019comparative}), i.e., how accurately a method is able to
correctly identify the genes that are differentially expressed. In the
multi-condition setting where the gene might show differences in
expression in one or more conditions, the corresponding question is
how well the method can correctly identify the genes that are
differentially expressed in {\em at least one condition} (Inference
Aim 1). However, it is also of interest to identify the specific
condition, or conditions, that give rise to differences in expression;
in other words, identifying the {\em gene-condition pairs} with
differences in expression (Inference Aim 2). We therefore evaluated
the performance of the methods in achieving both of these aims. We
also assessed accuracy of the expression difference estimates.

\subsection{Simulation design}
\label{sec:design}

We simulated data sets by applying ``binomial thinning''
\citep{gerard2020data} to the scRNA-seq data set described in 
Section \ref{sec:application} and in more detail in the Appendix. 
Binomial thinning is a technique intended to preserve as much 
as possible the properties of real scRNA-seq data.

The original scRNA-seq data contained UMI counts in different cell
subpopulations, but to simplify the simulations we focused on the
largest subpopulation, the B cells (Supplementary
Table~\ref{table:cytokines-data-summary}). 
For our simulations, we used cells from 25 of the 45 treatment conditions.
Note that the amount of expression data was fairly
even across all conditions (Supplementary Data).

Like most scRNA-seq data sets, the UMI counts were sparse; 93\% of the
UMI counts were zero. Also, like most scRNA-seq data sets, the average
expression rates of the genes varied greatly; the top 1\% of
genes by expression level had a median UMI count of 4.4 per cell,
whereas the median UMI count per cell across all genes was 0.02. The
sequencing depths (total UMI counts) also varied widely across cells;
they ranged from 500 to 20,000, with a median of 1,800. These aspects are
handled naturally by the Poisson mash model and other Poisson-based
models, whereas other models (e.g., limma, MAST) typically require
careful preprocessing (log-transformation and normalization) of the
data to account for these features.

From these data, we generated a smaller data set and a larger data
set.  For the smaller data set, we selected $N = \mbox{2,096}$ cells
uniformly at random from the 25 treatment conditions. For the larger
data set, we randomly selected $N = \mbox{15,705}$ cells from the same
25 conditions. In both data sets, we filtered out genes expressed in
fewer than 25 cells. After this filtering step, the small data set
contained $J = \mbox{8,358}$ genes and the large data set contained $J
= \mbox{10,691}$ genes.

Next, starting with either the small or large data set, we took the
following steps to simulate data sets.

First, we generated ``null'' data by randomly shuffling the treatment
labels among the cells. This eliminated systematic differences in
expression among treatments. Importantly, we shuffled the treatment
labels in the same way for all genes. This preserves correlations
among the genes that could confound the detection of expression
differences; this same approach was used in
\cite{gerard2020empirical}. Note that any random effects that might
have been specific to some treatment conditions were evened out across
conditions by the random shuffling procedure.

After creating the null data, we used binomial thinning to add treatment
effects to some genes, independently of the unwanted variation in the
data. 
Once a treatment effect was chosen, binomial thinning involved
simulating new counts from a binomial distribution {\em conditioned on
  the original counts}. To illustrate, consider the simpler case of two
conditions. In this case, simulating an increase in expression of gene
$j$ in condition 2 relative to condition 1 --- that is, $\beta_{j2} >
0$ --- involved simulating a ``thinned'' count for each cell $i$ in
condition 1 as $y_{ji}^{\mathrm{new}} \sim \mathrm{Binomial}(y_{ji},
e^{-\beta_{j2}})$. \cite{gerard2020data} explains how this idea is
extended to more than two conditions.

In the smaller data sets, we added treatment effects in this way to
600 out of the 8,358 genes chosen uniformly at random from the subset
of genes with a total UMI count of at least 200. In the larger data
sets, we added treatment effects to 1,000 genes chosen uniformly at
random among the genes with a total UMI count of at least 229.
For each selected gene $j$, we simulated the effect vector as
$\bm{\beta}_j = (\beta_{j1}, \ldots, \beta_{jR})'$ for the $R = 25$
treatment conditions from ${\bm\beta}_j = a_j w_j \bm{u}_j$, where the
sign $a_j \in \{-1, +1\}$ was $-1$ or $+1$ with equal probability,
$w_j$ was drawn uniformly at random from $[\log 1.5, \log 5]$, and
${\bm u}_j \in \mathbb{R}^{25}$ was drawn uniformly at random from
three sharing patterns (Supplementary Figure \ref{fig:design}): 
the first sharing pattern simulated the
situation in which only a single treatment condition (condition 21)
had differences in expression; and the second and third sharing
patterns simulated the situation in which the differences in
expression were shared among subsets of treatments (conditions 9--11
and 12--19, respectively). Note that several of the conditions ---
conditions 1--8, 20 and 22--25 --- were ``null'' conditions in that
there were no gene expression differences in {\em any} of the
simulations for these conditions. We arbitrarily treated the first
treatment condition as the control, and LFCs were defined with respect
to this control condition. Since the the largest $u_{jr}$ was always
1, this produced fold changes that were at most 5 in magnitude (or the
LFCs that were at most $\log_2 5 \approx 2.3$ in magnitude). Under
these settings, many of the true effects $\beta_{jr}$ were small
enough that there would be a benefit to pooling information across
multiple conditions.

We repeated this procedure 20 times for the large data set and another
20 times for the small data set to produce a total of 40 simulated
data sets.

\subsection{Methods compared}

We compared two variants of Poisson mash --- Poisson mash with and
without the ``RUV'' enhancement --- and several alternative methods
that have been published and have good software implementations. (Both
variants of Poisson mash included the ``random effect'' enhancement.)

The method most comparable to Poisson mash is mash
\citep{urbut2019flexible}.
mash takes as input a $J \times R$ matrix of condition-level
expression estimates and another $J \times R$ matrix containing the
standard errors of these estimates. We ran limma
\citep{smyth2004linear} to generate these matrices.

To assess the benefits of mash and Poisson mash over methods that
cannot exploit sharing of expression differences across conditions, we
also compared with standard DE analysis methods. Specifically, we
included three DE analysis methods --- limma \citep{smyth2004linear},
edgeR \citep{robinson2010edger} and MAST \citep{finak2015mast} --- which
were among the best-performing methods in the \cite{soneson2018bias}
benchmarking study.
These three methods (as well as other widely used DE analysis methods)
effectively make the assumption that {\em expression differences are
  independent across conditions.} 

We also included the nonparametric Kruskal-Wallis test in our
comparisons \citep{kruskal1952use}. Although the rank-based
Kruskal-Wallis test is not frequently used for DE analysis, it is
often used in other settings to detect differences among multiple
groups, and therefore it is natural to compare mash and Poisson mash
with the Kruskal-Wallis test. A benefit of the Kruskal-Wallis test is
that it is nonparametric and therefore should be less sensitive to
modeling assumptions. On the other hand, a disadvantage of the
Kruskal-Wallis test is that it does not provide condition-level
results, therefore we only used it for Inference Aim 1.

An important yet under-appreciated aspect of DE analysis of scRNA-seq
data is that accounting for unwanted variation can substantially
improve accuracy. (Although this aspect seems to have been neglected
from many benchmarking studies, several methods have been developed to
fill this need; e.g., \citealt{ruvseq, svaseq, gerard2020empirical}.)
Therefore, to assess the benefits of accounting for unwanted variation, 
we compared Poisson mash with and without the additional terms 
capturing unwanted variation.

See Appendix \ref{sec:methods-details} for details on
how the methods were applied to the simulated data sets.

A fundamental difference between Poisson mash and the other methods is
that Poisson mash works with the aggregated (``pseudobulk'') data,
${\bm X}$, whereas the other methods work with the cell-level data
${\bm Y}$, or a transformed version of ${\bm Y}$. The benefits and
drawbacks of a pseudobulk analysis have been studied elsewhere
\citep{lun2017overcoming, ahlmann2020glmgampoi, muscat,
  squair2021confronting},
and assessing and understanding them remains an active research
question. Since our principal aim is to evaluate Poisson mash and
compare against alternatives, and not to study the benefits of
pseudobulk analysis, we have tried to design the simulations so that
there should be no particular benefit to analyzing ${\bm X}$ instead
of ${\bm Y}$. However, since Poisson mash and the other methods are
based on different models, and are estimating different parameters, it
is difficult to separate the benefits of analyzing ${\bm X}$ vs. ${\bm
  Y}$ from other aspects.

\subsection{Performance evaluation}

To evaluate the methods in the simulations, we used the following
performance metrics.

For Inference Aim 1 (``detect differentially expressed genes''), the
true differentially expressed genes were defined simply as the genes
$j$ for which $\beta_{jr} \neq 0$ in at least one condition
$r$. (Recall, we have defined the LFC with respect to the first
condition, and in the simulated data sets $\beta_{j1}$ was always
zero.) We then summarized performance using power and false discovery
rate (FDR) as the {\em p}-value or {\em lfsr} threshold for reporting
DE genes was varied from 0 to 1. Power and false discovery rate (FDR)
were calculated as $\mathrm{FDR} \coloneqq
\frac{\mathrm{FP}}{\mathrm{TP + FP}}$ and $\mathrm{power} \coloneqq
\frac{\mathrm{TP}}{\mathrm{TP + FN}}$, where FP, TP, FN, TN denote the
number of false positives, true positives, false negatives and true
negatives, respectively.

For Inference Aim 2, we evaluated detection and estimation of
expression differences, again relative to the control condition. We
defined the true differentially expressed gene-condition pairs as all
pairs $(j, r)$, $r > 1$, such that $\beta_{jr} \neq 0$, then we
calculated power and FDR for this task. Following
\cite{urbut2019flexible}, we only considered a gene-condition pair to
be a true positive if the {\em p}-value or {\em lfsr} met the
threshold and if the sign of $\beta_{jr}$ was correctly estimated.

To evaluate the accuracy of the LFC estimates, we calculated the root
mean squared error $\mathrm{RMSE} \coloneqq \sqrt{\mathrm{MSE}}$ for
a specified subset of genes $G$ as
\begin{equation*}
\mathrm{MSE} \coloneqq \;
\frac{1}{|G| \times (R-1)} \sum_{j \,\in\, G} \sum_{r=2}^{R}
\big[(\hat{\beta}_{jr} - \hat{\beta}_{j1}) - 
(\beta_{jr} - \beta_{j1}) \big]^2.
\end{equation*}

\subsection{Simulation results}

\begin{figure}
\centering
\includegraphics[width=\textwidth]{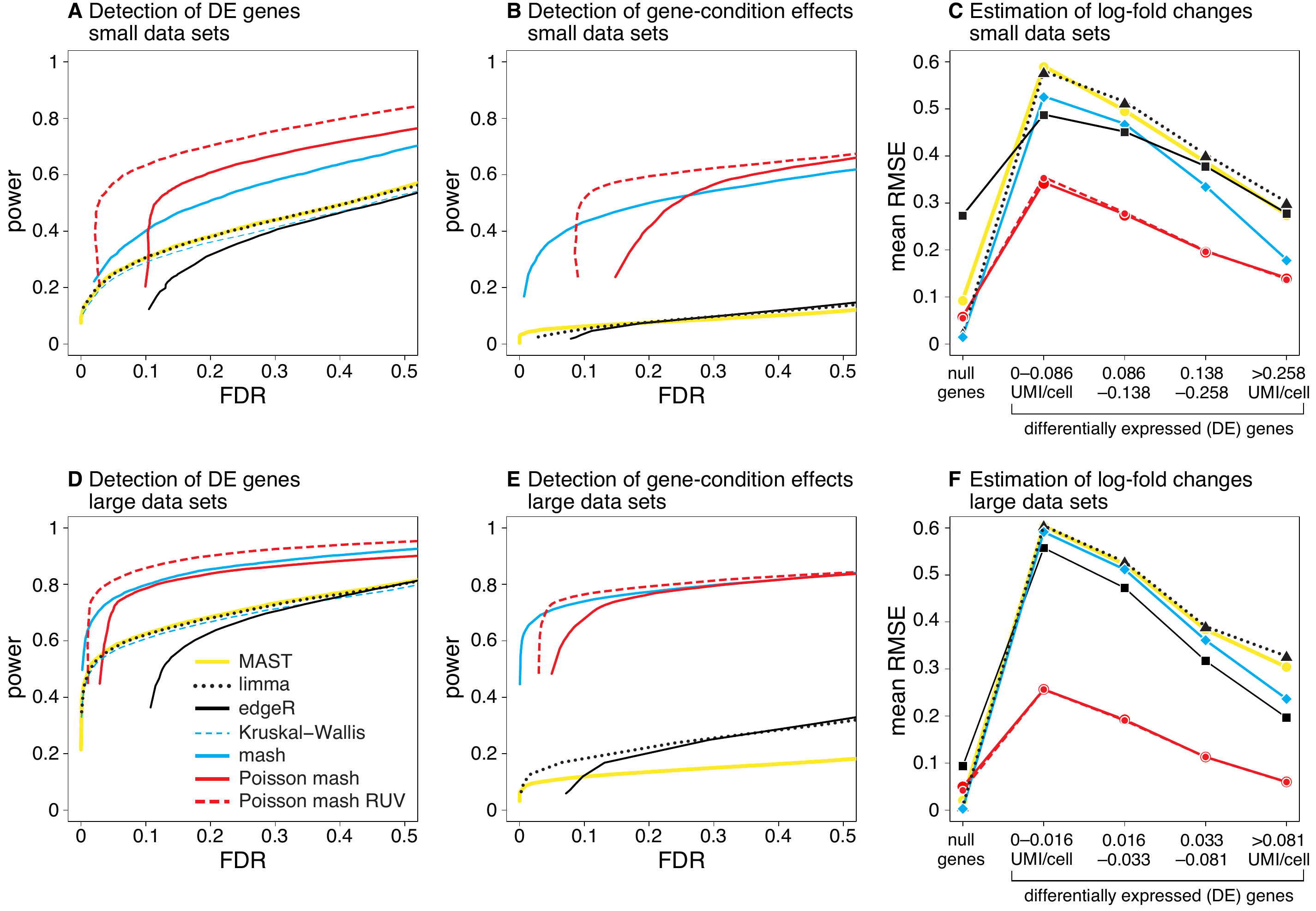}
\caption{\rm Evaluation of DE analysis methods in the small (top row)
  and large (bottom row) simulated data sets. FDR and power were
  calculated for all genes (A, D) and for all gene-condition
  pairs (B, E) in the 20 simulations by varying a {\em p}-value or
  {\em lfsr} threshold from 0 to 1. Panels C and F summarize LFC
  estimation accuracy by the RMSE, averaged over 20 simulations. RMSE
  was calculated in non-overlapping sets of genes: ``null'' genes
  (genes in which there were no differences in expression in all
  conditions); and DE genes grouped by expression level (counts of
  UMIs per cell).
  Note that the Kruskal-Wallis method was only included in A
  and D because it does not provide condition-level inferences.}
\label{fig:sims_main}
\end{figure}

\begin{figure}
\centering
\includegraphics[width=\textwidth]{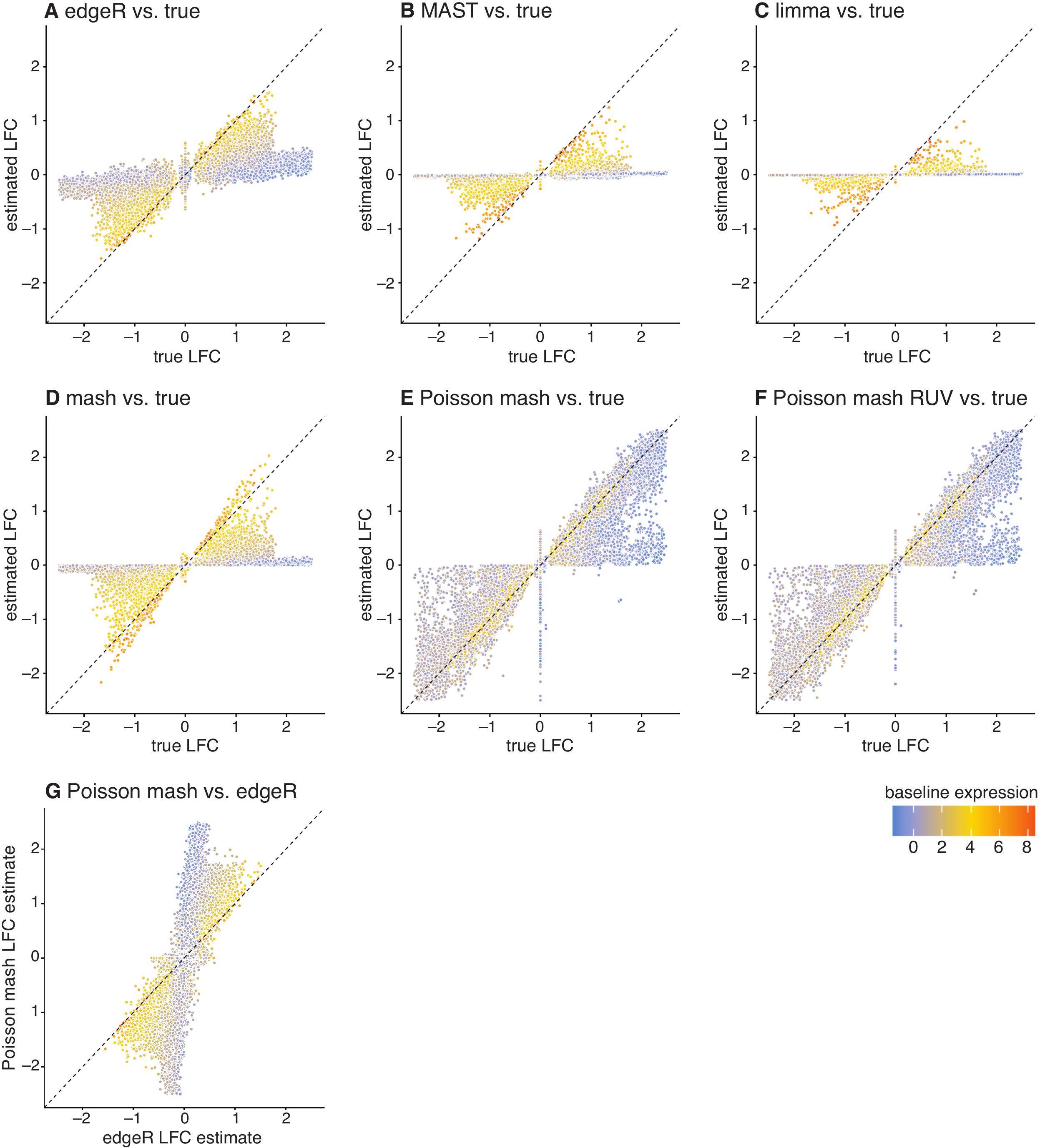}
\caption{\rm More detailed evaluation of LFC estimation accuracy. Each
  scatterplot shows the true LFC against the estimated LFC (in log
  base-2), except for the scatterplot at the bottom, which compares
  the edgeR and Poisson mash estimates. Each point depicts a result
  for a gene-condition pair $(j, r)$. DE genes are shown only from 5
  of the large simulated data sets, so each scatterplot shows results
  for $5 \times \mbox{1,000} \times 24 = \mbox{120,000}$
  gene-condition pairs. Note that most true $\beta_{jr}$ values are
  zero, even for DE genes, because a minority of conditions have
  differences in expression. The Kruskal-Wallis method is not included
  in these results because it does not provide condition-level
  estimates. All points are colored by the baseline expression level,
  $\mu_j$, that was estimated by Poisson mash RUV.}
\label{fig:lfc_scatterplots}
\end{figure}

First, we compared the methods' ability to detect DE genes and
condition-level expression differences. These comparisons are
summarized in Figure \ref{fig:sims_main}, Panels A, B, D and E. These
plots show power and FDR for each method as the {\em p}-value
threshold (for edgeR, MAST, limma and Kruskal-Wallis) or {\em lfsr}
threshold (for mash and Poisson mash) is varied. As expected, the
methods typically performed better in the larger data sets.
Compared to edgeR, MAST, limma and the Kruskal-Wallis test, which do
not model effect sharing across conditions, mash, Poisson mash and
Poisson mash RUV achieved much greater power in both the small and
large data sets.  (The Kruskal-Wallis does not provide condition-level
inferences so cannot compete this task.) The performance gains are
particularly striking for detecting condition-level expression changes
(B, E).

Next, we compared the methods' ability to estimate the condition-level
expression differences. Summarizing the estimation accuracy of LFCs by
gene expression level (Figure \ref{fig:sims_main}, Panels C and F),
Poisson mash and Poisson mash RUV had by far the best accuracy at all
expression levels, with the greatest gains for genes at lower
expression levels.

To understand these performance gains in greater detail, in Figure
\ref{fig:lfc_scatterplots} we compared the true LFCs $\beta_{jr} -
\beta_{j1}$ versus the estimated LFCs $\hat{\beta}_{jr} -
\hat{\beta}_{j1}$. MAST and limma usually underestimated the LFCs,
except for the most highly expressed genes. Since mash was provided
with the limma estimates as input, mash also suffered from the same
underestimation issue, although it was able to improve accuracy for
effects that were shared across conditions. Similar to mash, Poisson
mash also improved accuracy of the effect estimates but, unlike mash,
Poisson mash was not disadvantaged by the poor initial estimates
provided by limma. As a result, Poisson mash was often very accurate
even for lowly expressed genes.

Since limma and MAST and, by extension, mash, work with
log-transformed counts, whereas Poisson mash works with the ``raw''
counts, a key question is to what extent we should attribute these
improvements to (i) combining of information across conditions using
the flexible mash priors and (ii) to the use of a Poisson model of the
counts instead of a Gaussian model of the log-transformed counts. (The
log-transformation introduces biases in the LFC estimates, and this
bias remains regardless of the choice of pseudocount used in the
log-transformation; see Supplementary Figure \ref{fig:figure4}.) 
To help answer this question, we
compared Poisson mash to edgeR, which is also based on a Poisson model
of the counts. The results of running edgeR show that while the
underestimation problem is not quite as severe in edgeR, the overall
pattern is similar to MAST and limma; for all three methods, the LFCs
were estimated accurately mainly for highly expressed genes only. Only
after combining the Poisson model with informative mash priors did we
improve accuracy in lowly expressed genes as well.

Despite the fact that Poisson mash was much more accurate than mash,
surprisingly this improvement did not necessarily translate to an
improvement in power, particularly at lower false discovery rates ---
compare mash to Poisson mash in Panels A, B, D and E in Figure
\ref{fig:sims_main}. Poisson mash provides improvements in power {\em
  only after} including additional factors in the model to account for
unwanted variation, i.e., Poisson mash RUV. Poisson mash RUV does not
noticeably increase accuracy of the LFC estimates over Poisson mash
(Figure \ref{fig:sims_main}, Panels C and F), but it substantially
improves the ability of Poisson mash to detect DE genes and
condition-level expression differences (Figure \ref{fig:sims_main},
Panels A, B, D, E). Poisson mash RUV is only worse than mash at low
FDRs. It is possible that improvements to estimation of the
overdispersion parameters $\psi_j^2$ and unwanted variation
coefficients $f_{jd}$ --- say, by adaptively shrinking these
parameters jointly across all genes $j$ \citep{love2014moderated,
  robinson2010edger} --- may close the gap between mash and Poisson
mash RUV at low FDRs. Despite this one limitation, the simulations
show that Poisson mash RUV provides the best overall combination of
(i) strong performance in detecting expression differences and (ii)
accurate estimation of these differences.

\subsection{Simulations with fewer conditions}
\label{small-R}

Although we expect Poisson mash to be most beneficial in
multi-condition data sets with many conditions, it is also of interest
to know whether Poisson mash can cope with a smaller number of
conditions. Therefore, we performed additional simulations with $R =
6$ and $R = 12$ conditions. We simulated 20 $R = 6$ data sets and
another 20 $R = 12$ data sets, similar to above, with the following
changes: the $R = 6$ data sets had $J = \mbox{9,160}$ genes and $N =
\mbox{3,898}$ cells; the $R = 12$ data sets had $J = \mbox{10,003}$
genes and $N = \mbox{7,960}$ cells; in each data set, 1,000 of the
genes were chosen uniformly at random to have treatment effects, and
the non-null effects were generated from one of two effect-sharing
patterns: a condition-specific effect and an effect shared among
multiple conditions (Supplementary Figure \ref{fig:design}).

\begin{figure}
\centering
\includegraphics[width=\textwidth]{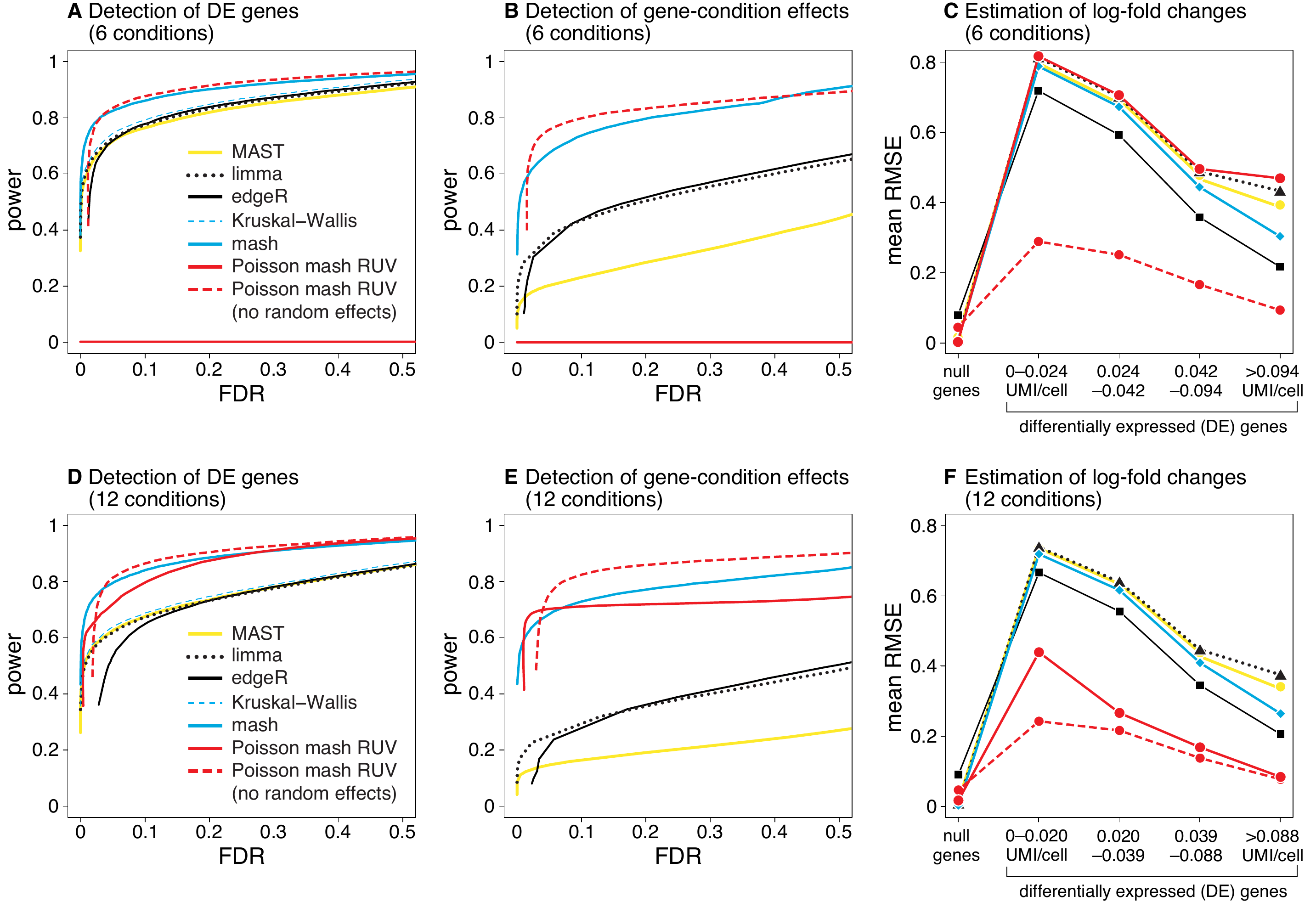}
\caption{\rm Evaluation of DE analysis methods in simulated data sets
  with fewer conditions. FDR and power were calculated for all genes
  (A, D) and for all gene-condition pairs (B, E) in the 20
  simulations by varying a {\em p}-value or {\em lfsr} threshold from
  0 to 1. Panels C and F summarize LFC estimation accuracy by the
  RMSE, averaged over 20 simulations. RMSE was calculated in
  non-overlapping groups of genes: ``null'' genes (genes in which
  there were no differences in expression in all conditions); and DE
  genes grouped by expression level (counts of UMIs per cell).
  The results of the Kruskal-Wallis method are only included in
  Panels A and D because this method does not provide condition-level
  inferences.}
\label{fig:sims_additional}
\end{figure}

The results of these simulations are summarized in Figure
\ref{fig:sims_additional}. While Poisson mash RUV still maintained
good performance with 12 conditions, it performed poorly in the data
sets with only 6 conditions. In particular, at a {\em minimum lfsr}
threshold of 0.05, Poisson mash RUV identified an average of 571 DE
genes in the $R=12$ data sets, but an average of only 0.2 DE genes in
the $R=6$ data sets. Upon closer examination, in the data sets with
only 6 conditions Poisson mash RUV had much more difficulty
distinguishing  condition-specific expression differences
$\beta_{jr}$ from random effects $\eta_{jr}$; in particular, Poisson
mash RUV tended to attribute observed variability across conditions to
random effects. To verify that this was indeed the case,
we re-ran Poisson mash RUV without the random effects.
This resulted in much better performance in all tasks (Figure
\ref{fig:sims_additional}).
Recall, Poisson mash is applied to the condition-level aggregated
data, so $\beta_{jr}$ and $\eta_{jr}$ are only identifiable by having
priors with different cross-condition covariance structures (Section
\ref{sec:random_effects}).  In other words, Poisson mash relies solely
on the patterns of observed variability across conditions to separate
the two types of effects, and is therefore expected to work better
when we have data from more conditions.

While removing the random effect term was able to rescue the
performance of Poisson mash RUV for $R=6$ in this particular
simulation, more generally it could hurt performance since random
effects might exist and need to be accounted for. A better
solution to this problem would be to collect multiple replicates
within each condition, and to extend Poisson mash to allow for
replicates; this should allow reliable estimation of the random
effects by using the variability between the replicates in each condition
(Appendix \ref{sec:replicate}). In the absence of this, 
users should be aware of the difficulty of distinguishing treatment vs. random 
effects ($\beta_{jr}$ vs. $\eta_{jr}$) when analyzing data with few conditions
(say, $R<10$), and interpret results from Poisson mash or Poisson mash
RUV with this in mind.

\subsection{Bulk RNA-seq simulations with multiple replicates} 
\label{sec:bulk}

While the focus of this paper is single-cell data, Poisson mash can
also be applied to bulk data, in which the columns of ${\bm Y}$
represent samples or replicates rather than cells.  Therefore, we also
evaluated Poisson mash in simulated bulk RNA-seq data sets. We
simulated bulk RNA-seq data for $R = 12$ and $R = 25$ conditions, with
4 replicates per condition (for details see Appendix \ref{sec:bulk-sims}). 
Then we applied Poisson mash to the data matrix ${\bm X}$ obtained by 
summing RNA-seq counts across replicates from the same condition. 
We compared Poisson mash with mash and with two other methods 
(limma, edgeR) that are widely used to analyze bulk RNA-seq data. 
(Note that mash, edgeR and limma were applied to the sample-level 
data ${\bm Y}$, whereas Poisson mash was applied to the aggregated 
condition-level data ${\bm X}$. The methods were run on the bulk data 
sets in the same way as in the single-cell simulations.)

The results of these comparisons are shown in Supplementary Figure
\ref{fig:sims_bulk}. Similar to the results
on simulated single-cell data sets, Poisson mash provided considerable
gains in detecting differential expression and in LFC estimation
compared to limma and edgeR, which analyze the data from each
condition independently. (Note that we did not include Poisson mash RUV in this
experiment because we did not simulate these data with an ``unwanted
variation'' component.)  However, in contrast to the scRNA-seq
simulations, mash and Poisson mash performed similarly well in all
inference tasks; only in estimating the LFCs was Poisson mash
marginally more accurate than mash across all simulation settings and
gene expression levels (Supplementary Figure \ref{fig:sims_bulk},
Panels C and F). These results suggest
that the bias in LFC estimates introduced by log-transformation is
much less severe for bulk data; consider that the counts are much
larger and less sparse in bulk data. Based on these results, we
concluded that Poisson mash does not provide a clear advantage over
mash for bulk RNA-see data. Considering that mash is simpler and more
flexible than Poisson mash --- for example, mash can easily
incorporate replicates, whereas this feature is not yet
implemented in Poisson mash --- our general recommendation would be to
use mash for multi-condition, multi-replicate bulk RNA-seq data.

\section{Application to cytokine stimulation data}
\label{sec:application} 

To illustrate the potential for our methods to provide new insights 
in real applications, we analyzed
data from a study of the effects of cytokine stimulation on gene
expression. Single-cell RNA-seq data were collected from peripheral
blood mononuclear cells (PBMCs) in cytokine-injected mice. Experiment
protocols and data preparation steps are described in  
Appendix \ref{sec:protocols}. These data were deposited into the 
Gene Expression Omnibus (GEO) data repository 
under accession number GSE214633.

The prepared data set contained UMI counts for 14,853 genes from
141,962 cells in $R = 45$ conditions (44 cytokine treatments and one
control). Most of the cells (138,142) were assigned to one of 8 cell
types based on expression of known marker genes (Supplementary
Table \ref{table:cytokines-data-summary} and Supplementary Data). 
Cells collected in each cytokine 
treatment condition were derived from three mice, but the information
regarding which cells came from which mouse was not available. 
Therefore, DE analyses were performed as if there were just one 
replicate per condition. 

To study changes in gene expression induced by cytokine stimulation,
we applied Poisson mash RUV separately to the data for each of the 8
cell types. To account for unwanted variation, we estimated a $J
\times D$ matrix $\bm{F}$, with $D = 4$, from the cell-level data
${\bm Y}$ (see Appendix \ref{sec:poisson_mash_details}).

For the mash prior, we included ``canonical'' covariance matrices
capturing condition-specific expression differences, a ``null''
canonical covariance matrix for no effect in all conditions, and 7
data-driven covariance matrices. We
estimated the data-driven covariance matrices separately in each cell
type since the predominant sharing patterns might vary
across cell types. In total, each mash prior was a mixture of $K = 45
+ 1 + 7 = 53$ covariance matrices.
After fitting the Poisson mash RUV model with a mash prior, we
computed LFC estimates relative to the {\em median} across all
conditions, which served as a more robust estimate of the baseline
gene expression than the single control condition that could exhibit
higher or lower gene expressions by chance. 

Note that all of our analyses of these data used the Poisson mash RUV
model, but for brevity we sometimes refer to the method as ``Poisson
mash''.

\subsection{Poisson mash results for neutrophils} 
\label{sec:neutrophils} 

We begin with a more detailed investigation of the Poisson mash
results for a single cell type --- neutrophils --- before summarizing
the results for all 8 cell types. The prepared neutrophils data set
contained expression data for 8,543 genes in 13,362 cells;
among the 45 treatment conditions, the IL-1$\alpha$ treatment had the
most single-cell expression measurements (1,892 cells), and IL-13 had the
fewest (41 cells) (Supplementary Table \ref{table:cytokines-data-summary}
and Supplementary Data). Based on the {\em minimum lfsr} with a
threshold of $\alpha = 0.05$, Poisson mash identified 2,535 genes as
being differentially expressed in at least one of the 45 conditions.
In the presentation of the results, we focussed
on a subset of 27 of the 45 conditions, omitting the control condition
and the 17 chemokine treatments (chemokines are a distinctive subset
of the cytokines) because these treatments tended to show much less
differential expression and therefore the patterns of differential
expression were much less interesting to look at.

\begin{figure}
\centering
\includegraphics[width=\textwidth]{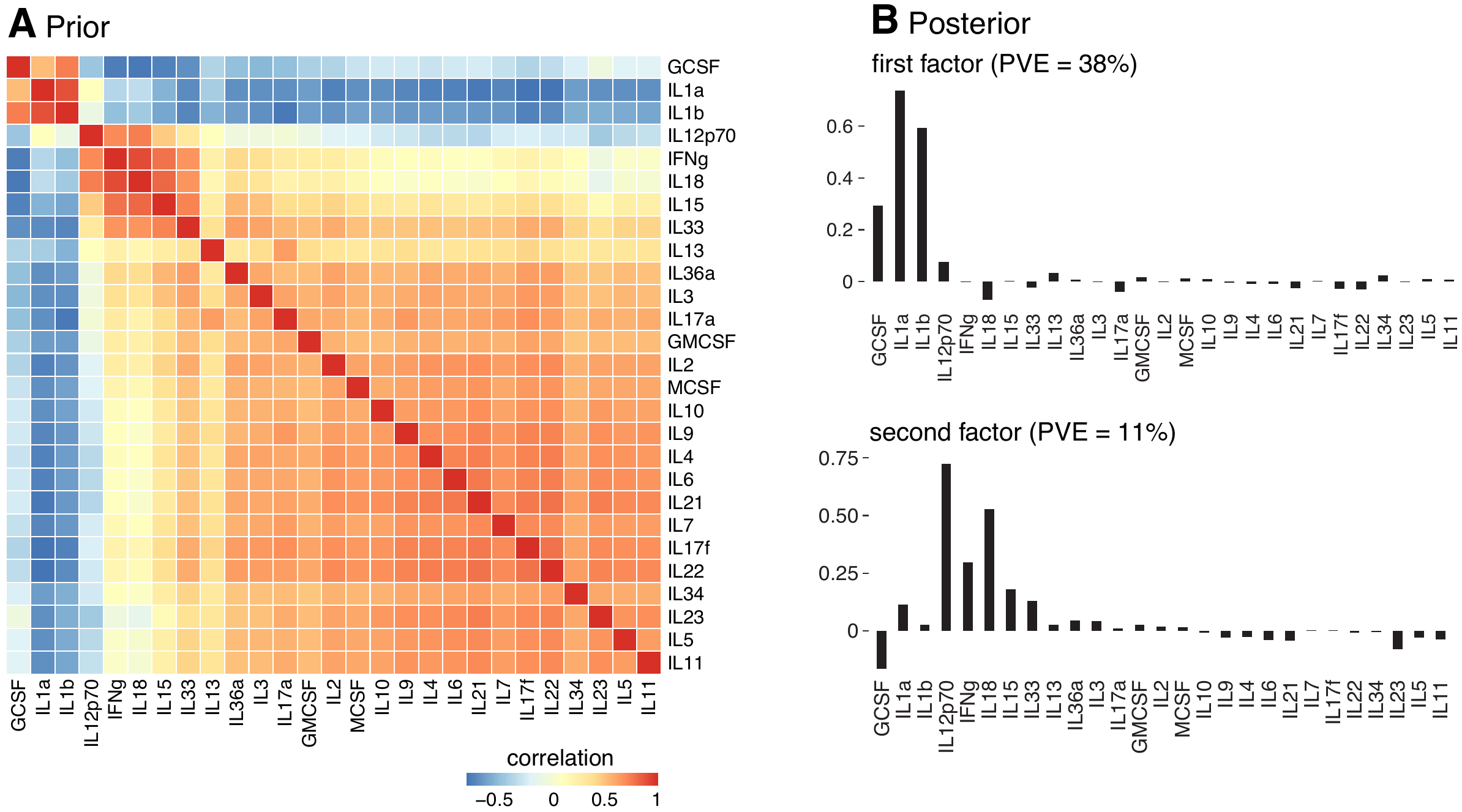}
\caption{\rm Patterns of differential expression across cytokine
  treatment conditions in the neutrophil data.
  The results shown here are for 27 of the 45 treatment
  conditions, omitting the control condition and the 17 chemokine
  treatments. Panel A shows correlations among cytokine treatments
  calculated from the top covariance matrix in the Poisson mash
  prior. The top covariance matrix is a data-driven covariance matrix,
  and accounts for 91\% of the total weight of the prior. Panel B
  shows the top two factors by PVE (proportion of variance explained)
  from a factor analysis of the Poisson mash RUV posterior mean LFC
  estimates.}
\label{fig:patterns_of_sharing}
\end{figure}

Poisson mash assigned most of the weight (91\%) in the mash prior to a
single covariance matrix (Figure
\ref{fig:patterns_of_sharing}A). This covariance matrix
captures particularly strong sharing of expression differences within
two subsets of cytokines: (1) IL-1$\alpha$, IL-1$\beta$, G-CSF;
and (2) IL-12 p70, IFN-$\gamma$, IL-18, IL-15, IL-33. There are
potential biological explanations for this structure. In the first
subset, IL-1$\alpha$, IL-1$\beta$ stimulate the same receptor, IL-1R1
\citep{dinarello2012treating}, and IL-1 has been shown to induce G-CSF
production \citep{altmeier20161}. We discuss this second subset below
in our analyses across cell types. To verify these sharing patterns,
we performed a factor analysis \citep{wang2021empirical} on the 
$\mbox{2,535} \times 27$ matrix of posterior mean LFCs. This factor analysis 
yields both a set of {\em factors}, each of which captures a pattern of 
sharing of DE effects among cytokines, and a corresponding set of {\em loadings}, 
which quantify how strongly each gene exhibits each pattern. The top two factors 
in this factor analysis (Figure \ref{fig:patterns_of_sharing}B) --- when ranked 
by proportion of variance explained --- capture similar patterns to those 
observed in the prior.

\begin{figure}
\centering
\includegraphics[width=\textwidth]{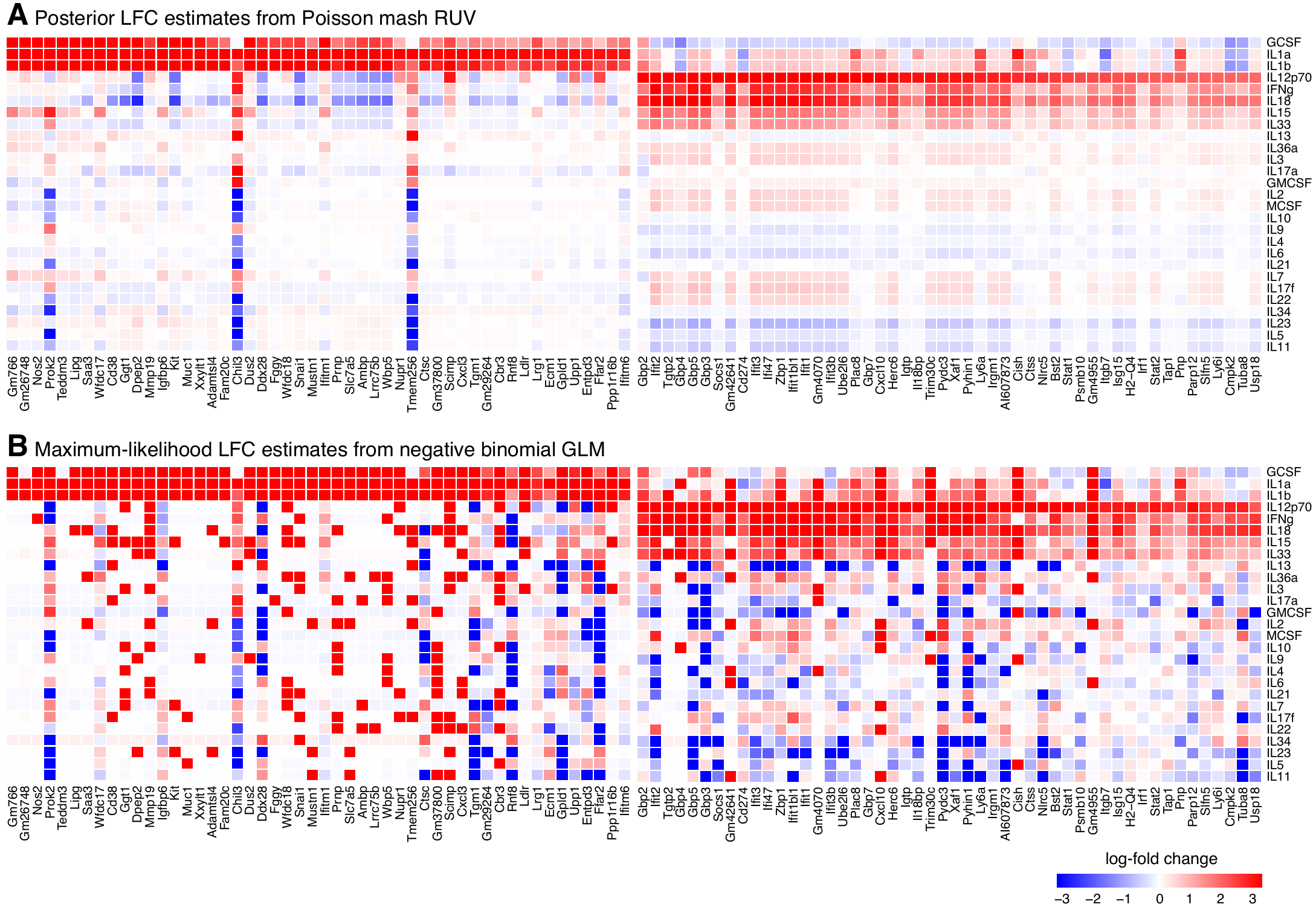}
\caption{\rm Multi-condition differential expression in neutrophils
  for selected genes as estimated by Poisson mash RUV (A) and a
  negative binomial GLM (B). The log-fold changes (in base-2 
  logarithm) are defined relative to the median expression level
  across cytokine treatments. Posterior mean estimates from Poisson
  mash RUV are shown in A, and maximum likelihood estimates from a
  negative binomial GLM model, fit separately to each gene, are shown
  in B. The selected genes are the 50 genes with the largest positive
  loadings in the first factor (Figure \ref{fig:patterns_of_sharing}B), 
  and another 50 genes with the largest positive loadings in the second
  factor (Figure \ref{fig:patterns_of_sharing}B). Results are shown for
  27 of the 45 conditions.}
\label{fig:LFC_comparison}
\end{figure}

To highlight the value of having Poisson mash learn these patterns
and then use these patterns to inform the estimation of DE effects,
we compared the Poisson mash posterior mean LFC estimates to the
maximum likelihood LFC estimates from a negative binomial (NB) 
GLM fit to the UMI counts, separately for each gene 
(Figure \ref{fig:LFC_comparison}; we used the {\tt glm.nb} 
function from the MASS R package). Reassuringly, the main patterns 
of shared DE effects are visually apparent in the maximum likelihood 
estimates, suggesting that our procedure is correctly identifying 
patterns that are present in the raw data. However, the maximum 
likelihood estimates can be noisy, especially for gene-condition 
pairs with small counts, and the stabilizing effect of the EB 
shrinkage estimation from Poisson mash is also visually apparent: 
the shrinkage estimation tends to ``denoise'' the LFC estimates 
by shrinking the smaller effects towards zero, and more generally 
making the estimates more concordant with the main patterns 
identified in the prior. Note that the effect estimates 
for the stronger effects are generally similar to the maximum 
likelihood estimates, illustrating that Poisson mash can 
regularize effect sizes adaptively and avoid over-shrinkage 
of true large effects.

For comparison, we also applied mash to these same
data. (Specifically, we applied mash to the LFC statistics returned by
running limma on the log-transformed UMI counts, as we did in the
simulations.) Consistent with the results of the simulations, mash
consistently produced posterior mean LFC estimates that were much
smaller in magnitude than Poisson mash (results not shown).

Finally, we assessed the goodness-of-fit of the Poisson mash
model fitted to the neutrophils data.
The result of the goodness-of-fit analysis is a set of 8,543 {\em
  p-}values, one for each gene, from a test of whether the data for
that gene could have been generated from the fitted model. The
histogram of the 8,543 {\em p-}values does not show a strong 
excess of zero or extremely small {\em p-}values 
(Supplementary Figure \ref{fig:gof}). Although the data do not
conform exactly to the Poisson mash model (the histogram is not
uniform), there are no strong ``outlier'' genes that show very
significant deviations from the fitted model.

\subsection{Patterns of cytokine response in eight cell types} 
\label{sec:interpret} 

To investigate patterns of cytokine response in all eight cell types,
we applied the procedure outlined above for neutrophils --- Poisson
mash, followed by a factor analysis of the Poisson mash posterior mean
LFC estimates --- separately to each of the eight cell types
(Supplementary Table \ref{table:cytokines-data-summary}). 
To interpret the factors, we used the
WebGestalt tool \citep{liao2019webgestalt} to identify Gene Ontology
(GO) gene sets \citep{go-2020} that were enriched in ``driving genes''
for each factor, which refer to the genes that exhibit the most
significant loadings ({\em lfsr} $< 0.001$) in that factor, then we
grouped factors together that shared enriched GO gene sets. As above,
these downstream analyses of the Poisson mash results were focussed on
the 27 conditions with the strongest overall differential expression
(omitting the control condition and the 17 chemokine treatments). See
Supplementary Figures \ref{fig:bcells}--\ref{fig:nk}
summarizing the results of the factor
analyses, and see Supplementary Table \ref{table:go-shared}
for details on the GO enrichment analyses.

Some of the factors captured patterns of cytokine response that were
relatively consistent across cell types, both in terms of the types of
cytokines involved and the enriched GO gene sets. These shared
factors are listed in Supplementary Table \ref{table:go-shared}.
For example, the cytokines IL-12 p70,
IL-18, IFN-$\gamma$, IL-15 and IL-33 --- the second subset we
mentioned earlier (Section \ref{sec:neutrophils}) --- were
identified in a factor in most cell types (except for dendritic 
cells and NK cells, where such a factor could still exist but 
may be undetected due to insufficient number of cells).
The driving genes for these factors were enriched for immune-related
GO terms such as response to interferon-beta, response to
interferon-gamma, and antigen processing and presentation. 
Consistent with this observation, previous studies have shown that
these cytokines induce the production of interferons
\citep{wojno2019immunobiology, jabri201515, dinarello2018overview}
which are responsible for the regulation and activation of the immune
system. For example, it has been shown that IL-18, initially named as
``IFN-$\gamma$ inducing factor'', can induce the production of
IFN-$\gamma$ by Th1 cells \citep{okamura1995cloning}. Furthermore,
IL-12 p40 deficient mice, which lack both IL-12 p40 and IL-12 p70, are
unable to control bacterial growth due to the absence of IFN-$\gamma$
production from splenocyte \citep{cooper1997interleukin}. In addition,
combination of IL-12 and IL-18 induce IFN-$\gamma$ production from B
cells in vitro \citep{yoshimoto1997interleukin}.

Other factors were more cell-type-specific, or at least more important
to one cell type. These cell-type-specific factors are listed in
Supplementary Table \ref{table:go-cell-type-specific}. 
For example, a factor capturing concordant
response to cytokines IL-15, IFN-$\gamma$, IL-23, IL-17A, IL-17F,
G-CSF, M-CSF appeared in CD4$^+$ T cells only. The driving genes were
enriched for GO terms such as response to wounding and fatty acid
metabolic process. Studies have shown that IL-23 is required for full
differentiation of effector Th17 cells \citep{mcgeachy2009interleukin}
which play a key role in tissue repair \citep{littman2010th17}, and
activation of mTOR signaling pathway \citep{chang2013myd88} which is
involved in regulation of lipid metabolism
\citep{shimobayashi2014making}. In addition, previous studies have
shown that IFN-$\gamma$ induces the expansion of IL-17-producing
CD4$^{+}$ T cells during mycobacterial infection
\citep{cruz2006cutting}, and intrathecal M-CSF expands CD4$^{+}$
regulatory T cells \citep{kuhn2021regulatory}, both of which are
actively involved in wound healing and tissue repair
\citep{boothby2020regulatory}.

We also identified factors in different cell types enriched for GO
terms related to cell cycle. This included factors capturing
concordant changes in IL-33 and IL-4 in CD4$^{+}$ T cells; IL-12 p70,
IL-17A, IL-17F, IL-1$\alpha$, IL-4, IL-11, IL-15, IL-33 in CD8$^{+}$ T
cells; and IL-17A, IL-15, IL-1$\beta$, IL-33, IL-34, IL-36$\alpha$ in
NK cells. Biologically, cell cycle is closely connected to cellular
proliferation, so these results are consistent with the fact that
different cytokines have been observed to induce proliferation in
different cell types. Furthermore, these results are consistent with
previous reports that IL-4, IL-33 and IL-15, respectively, induce the
proliferation of CD4$^+$ T cells, CD8$^+$ T cells and NK cells
\citep{zhu2002growth, dreis2019tissue, jabri201515}.

\section{Discussion}
\label{sec:discussion}

In this paper, we have introduced a flexible empirical Bayes method,
``Poisson mash'', for analyzing count data from many features (e.g.,
genes) observed in multiple conditions (e.g., treatments).  It
estimates differences in the underlying mean parameters while 
accounting for different patterns of effects across conditions.  Our
methods extend the mash framework \citep{urbut2019flexible} to
analysis of Poisson data.
Our work was motivated by differential expression analysis of
multi-condition scRNA-seq data, but the model is not specific to
scRNA-seq data and could also be useful for multivariate count data in
other settings.

Although differential expression analysis of scRNA-seq data has been
extensively studied, to the best of our knowledge our work is the
first that allows for arbitrary patterns of correlation across
multiple conditions, and uses these patterns to inform estimation and
testing of multi-condition differential expression. Our approach can
account for overdispersion of count data and unwanted
variation. Compared to existing methods, our approach achieves
substantial improvement in identifying and estimating
expression differences.
Another advantage of our empirical Bayes approach is that it yields
posterior distributions for the effects, which can be used to estimate
LFCs with respect to an arbitrary reference point such as the median
across conditions.
The posterior distributions can also be used to calculate the
probability that an expression difference is larger than $\delta$,
where $\delta > 0$ is some chosen LFC threshold.  For example,
$\delta$ might be an LFC considered to be ``biologically meaningful''
\citep{bochkina2007tail, treat}. This probability can be computed by
Monte Carlo simulation.

To fit the Poisson mash model, we used variational approximation
techniques to obtain tractable and scalable computations. This
approach involves approximating the true posterior of $\bm{\beta}_j$,
which does not have a closed form, by a mixture of multivariate
Gaussians. We made no other approximations. In particular, we did not
use a ``mean field'' approximation that makes conditional independence
assumptions. Such conditional independence assumptions often do not
hold in practice, and the mean field approximations are known to
generally underestimate uncertainty in posterior distributions
\citep{blei2017variational}. The approximate posteriors in Poisson
mash are expected to be accurate so long as a mixture of Gaussians is
a good fit for the true posterior.

A well-calibrated {\em lfsr} provides a condition-specific measure of
significance for DE testing. Our simulations showed that when true the LFCs
were sparse among conditions (i.e., non-null genes are differentially
expressed in only a small subset of conditions, corresponding to
sparse covariance matrices $\bm{U}_k$), the {\em lfsr} was
underestimated by both mash and Poisson mash RUV.  We found that this
was mostly due to the use of data-driven rank-1 covariance matrices in
the mash prior \citep{urbut2019flexible}. Rank-1 covariance matrices
reduce computational cost and are often more interpretable, but they
are also restrictive in that effects across conditions are forced to
lie within a rank-1 subspace of $\mathbb{R}^R$ and, as a result, this
produces a {\em lfsr} that cannot vary among conditions. This issue
could be partially addressed by adding a small positive number
$\epsilon$ to the diagonals of prior covariance matrices
$\bm{U}_k$. We found that choosing $\epsilon = 0.01$ worked well in
multiple settings, and perhaps a better choice of $\epsilon$ could be
chosen automatically from data. We reserve this question for future
work.  In addition, mis-estimation of the covariance matrices
$\bm{U}_k$ could also lead to {\em lfsr} calibration issues, and in
particular estimating non-sparse $\bm{U}_k$ when the true $\bm{U}_k$
is sparse.  Therefore, incorporating regularization in the estimation
of $\bm{U}_k$ to encourage sparsity could improve calibration of {\em
  lfsr}, and is a potential area of further investigation.

We took a ``pseudobulk'' approach that aggregates cells by condition
before performing the DE analysis.  This aggregation step inevitably
leads to a loss of information, preventing us from capturing more
complex patterns of DE between conditions
\citep{zhang2022ideas}. One could consider
instead modeling the cell-level counts. However, some previous studies
\citep{lun2017overcoming, squair2021confronting} have found that DE
methods that aggregate cells from each replicate perform better than
methods that model cell-level data and do not specifically account for
inter-replicate variation or intra-replicate dependence. Therefore, it
is not clear that the potential benefits of modeling cell-level counts
would be worth the added complexity of modeling and computation.

Another limitation of our current implementation is that it assumes
one replicate per condition. Incorporation of condition-specific
random effects, $\bm{\eta}_j$, into our model softens this
limitation, essentially by using variability across conditions to
serve as an imperfect replacement for within-condition
replicates. However, when multiple replicates are available in each
condition, it would clearly be advantageous to use them, and this could
improve the performance of Poisson mash when the number of conditions
is small. Our model could be extended to this case (see
Appendix \ref{sec:replicate}) but we have
not yet implemented this and we leave it to future work.

Although Poisson mash was motivated by multi-condition data,
in principle it should be straightforward to use Poisson mash to 
explore patterns of differential expression across multiple cell 
types. Since we have not applied Poisson mash for this purpose, 
it remains to be seen whether modeling shared effects is helpful 
for analyzing differences in expression across multiple cell types, 
but it seems plausible. One statistical issue that arises in such 
analyses is the ``double-dipping'' problem \citep{gao2022selective} 
which comes from performing differential expression analysis between 
groups that were, themselves, estimated from the same expression data. 
This double-dipping issue, however, is not specific to Poisson mash.

\section*{Acknowledgments}
We thank Abhishek Sarkar, Yuxin Zou and Jason Willwerscheid 
for their invaluable advice. We also thank the staff at the Research
Computing Center for providing the high-performance computing
resources used to implement the numerical experiments.

\clearpage

\sectionfont{\large}
\subsectionfont{\normalsize}

\begin{appendices}

\renewcommand{\thefigure}{S\arabic{figure}}
\renewcommand{\thetable}{S\arabic{table}}  
\setcounter{figure}{0}   
\setcounter{table}{0}   

\renewcommand\figurename{Supplementary~Figure}
\renewcommand\tablename{Supplementary~Table}

\section{Modeling unwanted variation at the condition level vs. at the cell
  level}
\label{sec:ruv_appendix}

Here we justify applying the Poisson mash RUV model
(\eqref{eqn:general_pm} in the main text) to aggregated data ${\bm
  X}$ when ${\bm F}$ is estimated from cell-level data ${\bm Y}$. See
the main text for definitions.

Consider the following model of the single-cell UMI counts,
\begin{equation}
\begin{aligned}
y_{ji} &\sim \mathrm{Pois}(\tilde{s}_i \tilde{\lambda}_{ji}) \\
\log\tilde\lambda_{ji} &= \mu_j + \sum_{r=1}^R \beta_{jr} \delta_{ri} +
\sum_{d=1}^D f_{jd} \alpha_{di},
\label{eq:general-pm-y}
\end{aligned}
\end{equation}
in which $\delta_{ri} = 1$ if cell $i$ is in condition $r$, otherwise
$\delta_{ri} = 0$, and $\alpha_{di}$ denotes the $d$th unobserved factor
causing unwanted variation in cell $i$. If we define the cell-level
size factors, $\tilde{s}_i$, by the total UMI counts,
\begin{equation*}
\tilde{s}_i = \sum_{j=1}^J y_{ji}
\end{equation*}
then it follows straightforwardly that the condition-level size
factors can be expressed as sums of the cell-level size factors:
\begin{equation*}
s_r \coloneqq \sum_{j=1}^J x_{jr}
= \sum_{i \,\in\, \mathcal{S}_r} \tilde{s}_i.
\end{equation*}
From this identity, the expected value of the aggregated count
$x_{jr}$ under model \eqref{eq:general-pm-y} is
\begin{align}
\E[x_{jr}] &=
\sum_{i \,\in\, \mathcal{S}_r} \E[y_{ji}] \nonumber \\
&=  \sum_{i \,\in\, \mathcal{S}_r} \tilde{s}_i \times
\textstyle
\exp \big(\mu_j + \beta_{jr} + \sum_{d=1}^D f_{jd} \alpha_{di} \big)
\nonumber \\
&= \exp(\mu_j + \beta_{jr}) \times
\sum_{i \,\in\, \mathcal{S}_r} \tilde{s}_i
   \textstyle \exp\big(\sum_{d=1}^D f_{jd} \alpha_{di}\big).
\end{align}
Next, by a Taylor series expansion, we have
\begin{align}
\E[x_{jr}] &\approx \exp(\mu_j + \beta_{jr}) \times
\sum_{i \,\in\, \mathcal{S}_r}
 \tilde{s}_i \textstyle \big(1 + \sum_{d=1}^D f_{jd} \alpha_{di} \big)
\nonumber \\
&= \exp(\mu_j + \beta_{jr}) \times
\textstyle \big\{ \sum_{i \,\in\, \mathcal{S}_r} \tilde{s}_i +
  \sum_{d=1}^D \sum_{i \,\in\, \mathcal{S}_r} \tilde{s}_i \alpha_{di} 
  f_{jd} \big\} \nonumber \\
&= s_r \times \exp (\mu_j + \beta_{jr})
\times \textstyle \big\{1 + \sum_{d=1}^D f_{jd} \rho_{rd} \big\},
\end{align}
in which define $\rho_{rd} \coloneqq \sum_{i \,\in\, \mathcal{S}_r}
\tilde{s}_i \alpha_{di} / s_r$. Finally, by a second Taylor series
expansion, we have
\begin{equation}
\E[x_{jr}] \approx s_r \times
\exp \bigg\{\mu_j + \beta_{jr} + \sum_{d=1}^D f_{jd} \rho_{rd} \bigg\}.
\label{eq:cell-vs-pseudobulk}
\end{equation}
This result suggests that we can include the additional term
$\sum_{d=1}^D f_{jd} \rho_{rd}$ in the Poisson mash model to
approximate the effect of the confounding variables on the aggregated
counts ${\bm X}$.

\section{Extension to multiple subgroups} 
\label{sec:subgroup}

So far, we have assumed a baseline expression level, $\mu_j$, that is
shared across all conditions. We have also implemented the setting in
which the $R$ conditions are subdivided into $M$ subgroups, and each
subgroup $\mathcal{T}_m \subseteq \{1, \ldots, R\}, m \in \{1, \ldots,
M\}$ has its own baseline $\mu_{jm}$.
For example, one may want to perform a multi-condition DE analysis in
$M$ cell types. The benefit of performing this analysis jointly for
multiple conditions and multiple cell types is that we can leverage
the sharing of expression differences across both conditions and cell
types.
The Poisson mash model for multiple subgroups is
\begin{equation}
\begin{aligned}
x_{jr} &\sim \mathrm{Pois} (s_r \lambda_{jr}), \\
\log \lambda_{jr} &= \mu_{jm(r)} + \beta_{jr} \\
\bm{\beta}_j &\sim \sum_{k=1}^K \sum_{l=1}^L
\pi_{kl} N_R(\bm{0}, w_l \bm{U}_k),
\end{aligned}
\end{equation}
in which $m(r)$ denotes the mapping from indices $r \in \{1, \ldots,
R\}$ to subgroups $m \in \{1, \ldots, M\}$ that corresponds to the
partitioning $\{\mathcal{T}_1, \dots, \mathcal{T}_M \}$.

\section{Extension to multiple replicates per condition (not implemented)}
\label{sec:replicate}

In the paper, we assumed only one replicate per condition. Here we
briefly describe the extension of Poisson mash to handle multiple
replicates per condition. This has the potential to improve estimation
of inter-replicate variances $\psi_j^2$ and therefore can help to
address identifiability issues (see Section \ref{sec:random_effects}
of the main text). {\em Please note that we have not yet implemented this 
extension, and we only describe this extension here to illustrate the 
potential of Poisson mash to be applied to data sets with
multiple replicates.}

Instead of aggregating the counts by condition, we aggregate
the counts by replicate; that is, $x_{jt}$ is the aggregated count for
gene $j$, replicate $t$, and $s_t$ is the size factor for replicate
$t$.

The Poisson mash RUV model for data with replicates is
\begin{equation}
\begin{aligned}
x_{jt} &\sim \mathrm{Pois}(s_t \lambda_{jt}) \\
\log\lambda_{jt} &= \mu_j + \beta_{jr} + \eta_{jt}
+ \sum_{d=1}^D f_{jd} \rho_{td},
\quad t \in \mathcal{T}_r.
\end{aligned}
\end{equation}
As before, $\bm{\beta}_j$ is a vector of length $R$, and is assigned
the mash prior \eqref{eq:mash_prior}; the random effects vector
$\bm{\eta}_j$ is a vector of length $N_T \coloneqq \sum_{r=1}^R
|\mathcal{T}_r|$ in which the individual elements are drawn independently
from the normal distribution with mean zero and variance $\psi_j^2$.
Also note that the matrix of unobserved factors, $\bm{\rho}$, is a
$N_T \times D$ matrix.

\section{Goodness-of-fit test}
\label{sec:goodness-of-fit}

Here we describe the test to assess goodness-of-fit for a Poisson mash
model. The description we give here is for the basic Poisson mash
model, but the same ideas easily extend to the Poisson mash models
with the various enhancements.

First, for gene $j$, we define the ``best ELBO'', denoted by
$\mathrm{ELBO}_j^{\star}$, as the best lower bound that can be
attained for a given setting of the model parameters:
\begin{align}
\label{eq:gof}
\mathrm{ELBO}_j^{\star}(\mu_j, \bm{\pi}, \bm{U}, \bm{x}_j)
\coloneqq \textstyle \max_{q_j}
\mathrm{ELBO}_j(q_j; \mu_j, {\bm\pi}, {\bm U}, {\bm x}_j).
\end{align}
Note we have modified the notation slightly here to make explicit the
dependence on the data, ${\bm x}_j$.  This quantity is lower bound to
the log-likelihood, $\log p(\bm{x}_j \mid \mu_j, \bm{\pi}, \bm{U})$,
and can be interpreted as an approximate goodness-of-fit measure for
data $\bm{x}_j$.

The goodness-of-fit test proceeds as follows. For gene $j$, we draw
i.i.d samples $\bm{x}_j^{(i)}$, $i = 1, \dots, n_s$, from the fitted
Poisson mash model, where $n_s$ is a (typically large) number of Monte
Carlo samples. Then we compute
\begin{equation}
\frac{1}{n_s} \sum_{i=1}^{n_s} \mathbbm{1}\big\{
\mathrm{ELBO}_j^{\star}(\mu_j, \bm{\pi}, \bm{U}, \bm{x}_j)
> \mathrm{ELBO}_j^{\star}(\mu_j, \bm{\pi}, \bm{U}, \bm{x}_j^{(i)}) \big\}.
\label{eq:pvalue}
\end{equation}
This quantity can be used like a {\em p-}value: if it is zero or close to
zero, it suggests that the Poisson mash model provides a poor fit for
$\bm{x}_j$; if the Poisson mash model fits the data well, the {\em
  p}-values across genes $j$ should roughly be uniformly distributed over
the $[0,1]$ interval.

\section{Model fitting algorithms}
\label{sec:fitting}

We now derive algorithms for fitting the Poisson mash RUV model to
data with multiple subgroups. For convenience, we summarize the model
here:
\begin{equation}
\begin{aligned}
x_{jr} &\sim \mathrm{Pois} (s_r \lambda_{jr}), \\
\log \lambda_{jr} &=
\mu_{jm(r)} + \beta_{jr}  + \eta_{jr} + \sum_{d=1}^D f_{jd} \rho_{rd} \\
\bm{\beta}_j &\sim \sum_{k=1}^K \sum_{l=1}^L
\pi_{kl} N_R(\bm{0}, w_l \bm{U}_k) \\
\bm{\eta}_j &\sim N_R(\bm{0}, \psi_j^2 \bm{I}_R).
\end{aligned}
\label{eq:poisson-mash-ruv-subgroups}
\end{equation}
The other variants of the Poisson mash model are special cases of this
model, and therefore the algorithms for Poisson mash RUV with multiple
subgroups can also be applied to the other variants of Poisson mash.

\subsection{Gaussian variational approximation}

In the main text, we described the posterior distribution for ${\bm
  \beta}_j$, and its Gaussian approximation, in the basic Poisson mash
model. Here we need to generalize the approach to the Poisson mash RUV
model. To derive the more general approach, we define
$\bm{\theta}_j \coloneqq \bm{\beta}_j + \bm{\eta}_j$, and we consider
the posterior distribution of ${\bm\theta}_j$, which is \\
\begin{equation}
p_{\mathrm{post}}(\bm{\theta} \mid {\bm x}_j, {\bm\Omega})
\propto p(\bm{x}_j \mid \bm{\theta}_j, {\bm\Omega})
\times \sum_{k=1}^K \sum_{l=1}^L \pi_{kl} N_R(
\bm{\theta}_j; \bm{0}, w_l \bm{U}_k + \psi_j^2 \bm{I}_R),
\end{equation}
where $\bm{\Omega} \coloneqq \{\bm{\mu}, \bm{\pi}, \bm{U}, \bm{\psi}^2,
\bm{\rho}\}$ denote the parameters to be estimated for the Poisson
mash RUV model.

Since this posterior does not have a closed form, we
approximate it by a mixture of multivariate normals,
\begin{equation}
q_j(\bm{\theta}_j) =
\sum_{k=1}^K \sum_{l=1}^L \zeta_{jkl}
N_R(\bm{\theta}_j; \bm{\varphi}_{jkl}, \bm{\Sigma}_{jkl}).
\label{eq:q-theta}
\end{equation}
Or, introducing the latent indicator variable ${\bm z}_j$ as we did in
the main text, the approximate posterior is
\begin{equation} 
q_j(\bm{\theta}_j, \bm{z}_j) = \prod_{k=1}^K \prod_{l=1}^L
\big\{\zeta_{jkl} 
N_R(\bm{\theta}_j; \bm{\varphi}_{jkl}, \bm{\Sigma}_{jkl})\big\}^{z_{jkl}}.
\end{equation}

\subsection{Evidence lower bound (ELBO)}
\label{sec:derivation}

We estimate the model parameters $\bm{\Omega}$ and the approximate
posteriors $q_1, \ldots, q_J$ by maximizing the ELBO:
\begin{equation}
\mathrm{ELBO}(q; \bm{\Omega}) =
\sum_{j=1}^J \mathrm{ELBO}_j(q_j; {\bm\Omega}),
\label{eq:elbo-poisson-mash-ruv}
\end{equation}
in which the ELBO for gene $j$ is
\begin{align}
\mathrm{ELBO}_j(q_j; {\bm\Omega}) &=
\log p({\bm x}_j \mid \bm{\mu}_j, {\bm\Omega}) -
D_{\mathrm{KL}}(q_j({\bm\theta}_j, {\bm z}_j) \,\lVert\,
p_{\mathrm{post}}({\bm\theta}_j, {\bm z}_j))
\\
&= \sum_{k=1}^K \sum_{l=1}^L \zeta_{jkl}
\big\{\log(\pi_{kl}/\zeta_{jkl})
+ \Psi_{jkl}({\bm\varphi}_{jkl}, \bm{\Sigma}_{jkl}; {\bm\Omega})\big\},
\end{align}
where
\begin{align} 
\Psi_{jkl}(\bm{\varphi}_{jkl}, \bm{\Sigma}_{jkl}; {\bm\Omega}) &\coloneqq
\mathbb{E}_q\big[\log p(\bm{x}_j \mid \bm{\theta}_j, {\bm\Omega})\big]
\nonumber \\
& \quad - D_{\mathrm{KL}}(
N_R(\bm{\theta}_j; \bm{\varphi}_{jkl}, \bm{\Sigma}_{jkl}) \,\lVert \,
N_R(\bm{\theta}_j; \bm{0}, w_l \bm{U}_k + \psi_j^2 \bm{I}_R)).
\end{align}
Expanding terms, this is
\begin{align*}
\Psi_{jkl}(\bm{\varphi}_{jkl}, \bm{\Sigma}_{jkl}; {\bm\Omega})
&= \sum_{r=1}^R \bigg\{x_{jr} \bigg(\log s_r + \mu_{jm(r)} 
+ \varphi_{jklr} + \sum_{d=1}^D f_{jd} \rho_{rd} \bigg) \bigg.
\\ 
& \qquad\qquad\qquad - s_r \exp\bigg(\mu_{jm(r)} 
+ \varphi_{jklr} +  {\textstyle\frac{1}{2}} \sigma_{jkl,rr}
+ \sum_{d=1}^D f_{jd} \rho_{rd} \bigg) \\
& \qquad\qquad\qquad - \log(x_{jr}!) \bigg\} \\
& \quad - D_{\mathrm{KL}}(
N_R(\bm{\theta}_j; \bm{\varphi}_{jkl}, \bm{\Sigma}_{jkl}) \,\lVert \,
N_R(\bm{\theta}_j; \bm{0}, w_l \bm{U}_k + \psi_j^2 \bm{I}_R)).
\end{align*}

\subsection{Model fitting algorithm with fixed prior covariance matrices}
\label{sec:maximization}

\begin{algorithm}[t]
\normalsize
  \caption{Variational Bayes for Poisson mash RUV
with fixed prior covariances.}
\label{alg:poisson_mash_ruv_fixed_U}
\begin{algorithmic}

\REQUIRE Count data $\bm{X}$ ($J \times R$ matrix), size factors
$\bm{s}$ (vector of length $R$), and additional covariates $\bm{F}$
capturing unwanted variation ($J \times D$ matrix).

\REQUIRE Prior covariances $\bm{U}_1, \ldots, \bm{U}_K$ ($R \times R$
matrices) and prior scaling factors $w_1, \ldots, w_L \geq 0$.

\REQUIRE Initial estimates of the model parameters, $\bm{\rho}, \bm{\pi}$.

\REQUIRE Initial estimates of the approximate posteriors, $\bm{\varphi},
\bm{\Sigma}, \bm{\zeta}$. 

\REQUIRE Convergence tolerances $\epsilon_{\mu}, \epsilon_{\psi},
\epsilon_{\Upsilon} \geq 0$.

\STATE $\bm{\Upsilon} = \bm{F} \bm{\rho}'$

\REPEAT

\STATE Perform all updates \eqref{eq:update-mu} and store the result in
${\bm\mu}^{\mathrm{new}}$.

\STATE Perform all updates \eqref{eq:update-psi} and store the result in
$({\bm\psi}^2)^{\mathrm{new}}$.

\STATE Compute $\argmax_{\bm\rho} \mathrm{ELBO}(q; {\bm \Omega})$ and
store the result in ${\bm\rho}$ (see
eq.~\ref{eq:update-rho}).

\STATE $\bm{\Upsilon}^{\mathrm{new}} = \bm{F} \bm{\rho}'$

\STATE $\mathcal{J}_{\mathrm{update}} \gets \{ j :
|(\psi_j^2)^{\mathrm{new}} - \psi_j^2| > \epsilon_{\psi}
\text{ or } 
\max_m |\mu^{\mathrm{new}}_{jm} - \mu_{jm}| > \epsilon_{\mu}
\text{ or }
\max_r |\Upsilon^{\mathrm{new}}_{jr} - \Upsilon_{jr}| > \epsilon_{\Upsilon}\}$

\FOR{$j = 1, \dots, J $}

\IF{$j \in \mathcal{J}_{\mathrm{update}}$}

\STATE $\bm{\mu}_{j} \gets \bm{\mu}_j^{\mathrm{new}}$

\STATE $\psi_j^2 \gets (\psi_j^2)^{\mathrm{new}}$

\STATE $\bm{\Upsilon}_j \gets \bm{\Upsilon}_j^{\mathrm{new}}$

\STATE For each $k, l$, compute $\argmax_{{\bm\varphi_{jkl}},
  {\bm\Sigma_{jkl}}} \mathrm{ELBO}(q; {\bm \Omega})$ by iterating
(\ref{eq:update-a}--\ref{eq:update-phi}).

\ENDIF

\ENDFOR

\STATE Perform all updates \eqref{eq:update-zeta} and store the result
  in ${\bm\zeta}$.

\STATE Perform all updates \eqref{eq:update-pi} and store the result in
  ${\bm\pi}$.

\UNTIL{$\mathcal{I} = \varnothing$.}

\RETURN parameter estimates $\bm{\mu}, \bm{\psi}^2, \bm{\rho},
\bm{\pi}$ and posterior estimates $\bm{\varphi}, \bm{\Sigma},
\bm{\zeta}$.

\end{algorithmic}
\end{algorithm}

We now describe a {\em coordinate ascent algorithm} for fitting the
Poisson mash RUV model with Gaussian variational approximations to the
posterior distributions of the unknowns ${\bm\theta}_j$. For now, we
assume that the prior covariances ${\bm U}_1, \ldots, {\bm U}_K$ are
known, and below we will describe an algorithm to estimate them.

The model fitting algorithm is summarized in
Algorithm~\ref{alg:poisson_mash_ruv_fixed_U}.

The most computationally intensive step in
Algorithm~\ref{alg:poisson_mash_ruv_fixed_U} is the updating of the
approximate posteriors for each gene $j$. To reduce computation
without greatly impacting accuracy, we selectively update the posteriors
based on the updates to the model parameters. The idea is that if the
parameters for gene $j$ did not change much, the posterior for gene $j$
will most likely not change much either. The genes $j$ to be updated
are stored in $\mathcal{J}_{\mathrm{update}}$.

\subsubsection{Coordinate ascent updates for the model parameters}

The update for the subgroup-specific baseline expression levels
${\bm\mu}$ maximizing the ELBO \eqref{eq:elbo-poisson-mash-ruv} with the
other parameters kept fixed is
\begin{align}
\mu_{jm}^{\mathrm{new}} &=
\log \big\{ {\textstyle\sum_{r \,\in\, \mathcal{T}_m} x_{jr}} \big\}
\nonumber \\
& \quad -
\log \big\{ {\textstyle \sum_{r \,\in\, \mathcal{T}_m} 
\sum_{k=1}^K \sum_{l=1}^L s_r \zeta_{jkl} \exp\big(\sum_{d=1}^D 
f_{jd} \rho_{rd} + \varphi_{jklr} + \frac{1}{2} \sigma_{jkl,rr}}\big) \big\}.
\label{eq:update-mu}
\end{align}

The update for ${\bm\psi}^2$ maximizing the ELBO
\eqref{eq:elbo-poisson-mash-ruv} with the other parameters unchanged is
\begin{align}
(\psi_j^2)^{\mathrm{new}} &=
\frac{1}{R} \mathbb{E}_{q_j}[\bm{\eta}_j' \bm{\eta}_j] \nonumber \\
&= \frac{1}{R} \sum_{k=1}^K \sum_{l=1}^L \zeta_{jkl}
\mathbb{E}_{q_j}[\bm{\eta}_j' \bm{\eta}_j \mid z_{jkl} = 1].
\label{eq:update-psi}
\end{align}
The expectations in this expression have a closed form because the
approximate posterior conditioned on the indicator variables
$z_{jkl}$ is multivariate normal.

The update for the mixture weights ${\bm\pi}$ maximizing the ELBO
\eqref{eq:elbo-poisson-mash-ruv} with the other parameters kept fixed
is simply
\begin{equation}
\label{eq:update-pi}
\pi_{kl}^{\mathrm{new}} = \frac{1}{J} \sum_{j=1}^J \zeta_{jkl}.
\end{equation}

The update for ${\bm\rho}$ does not have a closed form, so instead we
found the maximizer by Newton's method. Because the ELBO
\eqref{eq:elbo-poisson-mash-ruv} is separable over the $R$ rows of
${\bm\rho}$, we can compute the maximizer separately for each row of
${\bm\rho}$:
\begin{align}
\rho_{r1}^{\ast}, \ldots \rho_{rD}^{\ast} &\coloneqq
\argmax_{\rho_{r1}, \ldots, \rho_{rD}} \;
\mathrm{ELBO}(q; {\bm\Omega}) \nonumber \\
&= \argmax_{\rho_{r1}, \ldots, \rho_{rD}} \;
\bigg\{ \sum_{j=1}^J \sum_{d=1}^D x_{jr}  f_{jd} \rho_{rd} \nonumber \\
& \qquad - \sum_{j=1}^J \sum_{k=1}^K \sum_{l=1}^L
s_r \zeta_{jkl} \exp\big({\textstyle \mu_{jm(r)} +
\sum_{d=1}^D f_{jd} \rho_{rd} + \varphi_{jklr} +
\frac{1}{2} \sigma_{jkl,rr}}\big) \bigg\}.
\label{eq:update-rho}
\end{align}

\subsubsection{Coordinate ascent updates for the approximate posteriors}

The remaining part of the coordinate ascent algorithm are the updates
for the approximate posteriors $q_j$.

The mixture weights ${\bm\zeta}$ maximizing the ELBO---subject to the
constraints that they are non-negative and sum to 1 for each
$j$---have the following closed-form solution:
\begin{equation} 
\zeta_{jkl}^{\mathrm{new}} \propto \pi_{kl}
\exp\{{\Psi_{jkl}(\bm{\varphi}_{jkl}, \bm{\Sigma}_{jkl}; {\bm\Omega})}\}.
\label{eq:update-zeta}
\end{equation}

The updates to the posterior means and covariances do not have
closed-form solutions; instead we use the algorithm of
\cite{arridge2018variational}. This algorithm involves iterating the
following three updates until some convergence criterion is met or
until some predetermined upper bound on the number of iterations is
reached:
\begin{align}
  \label{eq:update-a}
a_{jklr} &=
s_r \exp\big\{{\textstyle \mu_{jm(r)} + \sum_{d=1}^D \rho_{rd} f_{jd} +
\varphi_{jklr} + \frac{1}{2} \sigma_{jkl,rr}}\big\} \\
\bm{\Sigma}_{jkl}^{\mathrm{new}} &=
\big((w_l \bm{U}_k + \psi_j^2 \bm{I}_R)^{-1} +
\mathrm{diag}(\bm{a}_{jkl})\big)^{-1}
\label{eq:update-sigma} \\
\bm{\varphi}_{jkl}^{\mathrm{new}} &= \bm{\varphi}_{jkl} -
\bm{\Sigma}_{jkl}^{\mathrm{new}} \left[ \bm{a}_{jkl} - \bm{x}_j +
\left(w_l \bm{U}_k + \psi_j^2 \bm{I}_R \right)^{-1}
\bm{\varphi}_{jkl} \right],
\label{eq:update-phi}
\end{align}
where $\bm{a}_{jkl} \coloneqq (a_{jkl1}, \ldots, a_{jklR})'$.

\subsection{Model fitting algorithm with unknown prior covariance matrices}
\label{ed}

Here we describe estimation of the ``data-driven'' covariance matrices
$\bm{U}_k$ using a simplified Poisson mash RUV model. These estimated
covariance matrices (possibly combined with other covariance matrices,
such as ``canonical'' covariance matrices) are then used to fit the
final Poisson mash RUV model
(Algorithm~\ref{alg:poisson_mash_ruv_fixed_U}).

To estimate the data-driven covariance matrices $\bm{U}_k$, we fit a
slightly simplified Poisson mash RUV model
\eqref{eq:poisson-mash-ruv-subgroups} without the mixture over the
different scales; specifically, with $L = 1$ and $\omega_1 = 1$.
Since $l$ is always 1, in the expressions we drop the ``$l$''
subscripts, and the mash prior for this simplified model is
\begin{equation} 
\bm{\beta}_j \sim \sum_{k=1}^K \pi_k N_R(\bm{0}, \bm{U}_k).
\label{eq:mash-prior-simplified}
\end{equation}
Similarly, without the mixture over the different scales, the
approximate posterior is
\begin{equation}
q_j(\bm{\theta}_j, \bm{z}_j) = \prod_{k=1}^K 
\big\{\zeta_{jk} 
N_R(\bm{\theta}_j; \bm{\varphi}_{jk}, \bm{\Sigma}_{jk})\big\}^{z_{jk}}.
\end{equation}
Not using a mixture of scales in the mash prior is
somewhat justified by the fact that we fit this model only to the
genes with the strongest effects (see
Section~\ref{sec:poisson_mash_details}).

\begin{algorithm}[t]
\normalsize
\caption{Variational Bayes for Poisson mash RUV
with simplified model and unknown prior covariances.}
\label{alg:poisson_mash_ruv_est_U}
\begin{algorithmic}

\REQUIRE Count data $\bm{X}$ ($J \times R$ matrix), size factors
$\bm{s}$ (vector of length $R$), and additional covariates $\bm{F}$
capturing unwanted variation ($J \times D$ matrix).

\REQUIRE Initial estimates of the prior covariance matrices ${\bm U}_1,
\ldots, {\bm U}_K$. The rank-1 covariance matrices are represented by
vectors ${\bm u}_k$ such that ${\bm U}_k = {\bm u}_k {\bm u}_k'$.

\REQUIRE Initial estimates of the other model parameters, $\bm{\mu},
\bm{\psi}^2, \bm{\rho}, \bm{\pi}$.

\REPEAT

\FOR{$j = 1, \dots, J$}

\STATE For each $k, l$, compute $\argmax_{{\bm\varphi_{jk}},
  {\bm\Sigma_{jk}}} \mathrm{ELBO}(q; {\bm \Omega})$ by iterating
(\ref{eq:update-a}--\ref{eq:update-phi}).

\STATE Update $\zeta_{j1}, \ldots, \zeta_{jK}$ using
  \eqref{eq:update-zeta}.

\ENDFOR

\FOR{$k = 1, \ldots, K$} 

\IF{${\bm U}_k$ is a rank-1 matrix}

\STATE Perform update \eqref{eq:update-U-rank-1} and store the result
  in ${\bm u}_k$.

\STATE ${\bm U}_k \leftarrow {\bm u}_k {\bm u}_k'$

\ELSE

\STATE Perform update \eqref{eq:update-U-full-rank} and store the
  result in ${\bm U}_k$.

\ENDIF

\ENDFOR   

\STATE Perform all updates \eqref{eq:update-mu} and store the result
  in ${\bm\mu}$.

\STATE Perform all updates \eqref{eq:update-psi} and store the result
  in ${\bm\psi}^2$.

\STATE Compute $\argmax_{\bm\rho} \mathrm{ELBO}(q; {\bm \Omega})$ and
store the result in ${\bm\rho}$ (see
eq.~\ref{eq:update-rho}).

\STATE Perform all updates \eqref{eq:update-pi} and store the result in
  ${\bm\pi}$.

\UNTIL{convergence criterion is met}

\RETURN ${\bm U}_1, \ldots, {\bm U}_K$
\end{algorithmic}
\end{algorithm} 

As before, we take a coordinate ascent approach to optimizing the
ELBO. Most of the coordinate ascent updates derived above can be
reused here by setting $L = 1, w_l = 1$. The only update we still need
to derive is the update for ${\bm U}_k$. Since some of these ${\bm
  U}_k$ may be rank-1 matrices, we also derive special updates for
rank-1 covariance matrices to take advantage of their special
properties. The coordinate ascent algorithm is summarized in
Algorithm~\ref{alg:poisson_mash_ruv_est_U}.

\subsubsection{Update for the full-rank prior covariance matrices}

For the simplified Poisson mash RUV model
\eqref{eq:mash-prior-simplified}, the ELBO, after dropping terms
that do not involve ${\bm U}_k$, is
\begin{align}
\mathrm{ELBO}(q; {\bm\Omega})
&= \sum_{j=1}^J \zeta_{jk} \mathbb{E}_{q_j}
\big[\log N_R(\bm{\beta}_j; \bm{0}, \bm{U}_k) \mid z_{jk} = 1 \big]
+ \mbox{constant} \nonumber \\
&= -\frac{1}{2} \sum_{j=1}^J \zeta_{jk}
\{\log|\bm{U}_k| + \mathrm{tr}(\bm{U}_k^{-1}
\mathbb{E}_{q_j}[\bm{\beta}_j \bm{\beta}_j' \mid z_{jk} = 1])\}
+ \mbox{constant.}
\end{align}
Therefore, we have the following closed-form update for the full-rank
covariance matrix ${\bm U}_k$:
\begin{equation}
\bm{U}_k^{\mathrm{new}} =
\frac{\sum_{j=1}^J \zeta_{jk}
      \mathbb{E}_{q_{jk}}[\bm{\beta}_j \bm{\beta}_j' \mid z_{jk} = 1]}
     {\sum_{j=1}^J \zeta_{jk}}.
\label{eq:update-U-full-rank}
\end{equation}
Using the fact that the distribution of ${\bm \beta}_j \mid z_{jk} =
1$ is multivariate normal under the variational approximation, the
expectations \eqref{eq:update-U-full-rank} in work out to
\begin{align} 
\mathbb{E}_{q_j}[\bm{\beta}_j\bm{\beta}_j' \mid z_{jk} = 1] &= 
(\bm{U}_k^{-1} + \bm{I}_R/\psi_j^2)^{-1}
\nonumber \\
& \quad + 
(\psi^2_j \bm{U}_k^{-1} + \bm{I}_R)^{-1}(\bm{\varphi}_{jk}
\bm{\varphi}_{jk}' + \bm{\Sigma}_{jk}) (\psi_j^2 \bm{U}_k^{-1}
+ \bm{I}_R)^{-1}.
\end{align}

\subsubsection{Update for the rank-1 prior covariance matrices}

If ${\bm U}_k$ is a rank-1 covariance matrix, the mash prior
conditioned on $z_{jk} = 1$, $\bm{\beta}_j \mid z_{jk} = 1 \sim
N_R(\bm{0}, \bm{U}_k)$, can be written as
\begin{equation}
\begin{aligned}
\bm{\beta}_j &= v_j \bm{u}_k \\
\quad v_j &\sim N(0,1),
\end{aligned}
\end{equation}
where ${\bm U}_k = {\bm u}_k {\bm u}_k'$. We can use this result to
write an update expression that is equivalent to the update above, but
has computational advantages; in particular, it avoids some expensive matrix
operations on $R \times R$ matrices such as matrix inversions.

After dropping terms that do not depend on ${\bm U}_k$, the ELBO is
\begin{align}
\mathrm{ELBO}(q; {\bm\Omega}) &=
\sum_{j=1}^J \zeta_{jk} \mathbb{E}_{q_j}[
\log N(\bm{\theta}_j; v_j \bm{u}_k, \psi_j^2 \bm{I}_R) \mid z_{jk} = 1] +
\mbox{constant} \nonumber \\
&= \sum_{j=1}^J \frac{\zeta_{jk} \bm{u}_k' 
  \mathbb{E}_{q_j}[v_j {\bm\theta}_j \mid z_{jk} = 1]}{\psi_j^2} 
- \sum_{j=1}^J \frac{\zeta_{jk} \bm{u}_k' \bm{u}_k
\mathbb{E}_{q_j}[v_j^2 \mid z_{jk} = 1]}{2\psi^2_j}
+ \mbox{constant.}
\label{eq:mash_prior1}
\end{align}
Solving for ${\bm u}_k$ yields the following update:
\begin{equation}
\bm{u}_k^{\mathrm{new}} = \frac{\sum_{j=1}^J \zeta_{jk}/\psi^2_j \times 
\mathbb{E}_{q_j}[v_j \bm{\theta}_j \mid z_{jk} = 1]}
{\sum_{j=1}^J \zeta_{jk}/\psi^2_j \times
\mathbb{E}_{q_j}[v_j^2 \mid z_{jk} = 1]},
\label{eq:update-U-rank-1}
\end{equation}
where
\begin{align}
\mathbb{E}_{q_j}[v_j^2] &=
\frac{\psi_j^2 + \bm{u}_k' \mathbb{E}_{q_j}[v_j \bm{\theta}_j]}
     {\bm{u}_k' \bm{u}_k + \psi_j^2} \label{eq:mash_prior3} \\
\mathbb{E}_{q_j}[v_j \bm{\theta}_j] &=
(\bm{\varphi}_{jk} \bm{\varphi}_{jk}' + \bm{\Sigma}_{jk}) \times
\frac{\bm{u}_k}{\bm{u}_k'\bm{u}_k + \psi_j^2}.
\label{eq:mash_prior4}
\end{align}

\section{Other simulation details}
\label{sec:methods-details}

Additional details about the simulations are given in this section.

\subsection{Details of the methods compared in the simulations}

Here we describe in detail how we ran each of the DE analysis methods
in the simulations.

\subsubsection{limma}
\label{sec:limma}

We used the {\em limma-trend} method \citep{law2014voom,
  phipson2016robust, smyth2004linear} implemented in the limma R
package (version 3.48.3; \citealt{limma}). This involved the following
steps. First, normalization factors were computed using the edgeR
function {\tt calcNormFactors} with the (default) ``trimmed mean of
M-values'' method \citep{robinson2010scaling}. Then we computed
log-normalized counts using the {\tt cpm} function from edgeR with
{\tt log = TRUE} and {\tt prior.count = 3}. Next, we fit a linear
model to the log-normalized counts using the limma function {\tt
  lmFit}, with conditions as variables; specifically, we chose the
``design matrix'' as {\tt model.matrix({\textasciitilde}0 +
  conditions)}, where {\tt conditions} was a categorical variable
encoding the assignment of cells to conditions. We reported
maximum-likelihood estimates of the LFCs. For DE detection, we used
the {\em p-}values from the moderated {\em F}-statistics which were
obtained by the function {\tt eBayes} with settings {\tt trend = TRUE}
and {\tt robust = TRUE} \citep{law2014voom, phipson2016robust}. Note
that, because the number of cells in each condition was quite large,
the moderated {\em F}-statistics rarely differed much from the
unmoderated {\em F}-statistics.
Still, the final {\em p-}values were based on the moderated {\em
  F}-statistics.

\subsubsection{MAST}
\label{sec:MAST}

The MAST R package (version 1.18.0; \citealt{mast}) implements the
hurdle GLM model proposed in \cite{finak2015mast}. Like limma, we fit
the MAST model to the log-normalized counts in which conditions were
included as variables in the GLM. The log-normalized counts were
computed in the same way as in limma, except that we set {\tt
  prior.count = 1}. As suggested by \cite{soneson2018bias}, we also
included the cellular detection rate (the fraction of genes detected
in each cell) as a covariate. We fit the GLM using the {\tt zlm}
function from MAST, keeping all default settings. For detecting DE
genes, we used the {\em p}-values from the likelihood-ratio test
(implemented by function {\tt lrTest} in MAST). For detecting and
estimating condition-level changes in expression, we called {\tt
  getLogFC} to get LFC estimates and variances of these
estimates. Two-tailed {\em p}-values were obtained from the {\em
  z}-scores, and these were used to identify differentially expressed
gene-condition pairs.

\subsubsection{Kruskal-Wallis test}
\label{sec:kw}

We performed the Kruskal-Wallis rank-sum test \citep{kruskal1952use}
on ${\bm Y}$ using function {\tt kruskal.test} from the stats R
package, in which the conditions defined the groups for the
Kruskal-Wallis test.

\subsubsection{edgeR}
\label{sec:edgeR}

We used the edgeR package (version 3.34.1; \citealt{edgeR-2012,
  robinson2010edger}) to fit a negative binomial GLM to the counts
${\bm Y}$, in which conditions were treated as variables in the
GLM. Following \cite{soneson2018bias}, we included the cellular
detection rate as an additional variable. We used edgeR functions {\tt
  estimateDisp} followed by {\tt glmQLFit} to fit the GLM, and tested
DE effects by the quasi-likelihood-ratio test using {\tt glmQLFTest}
\citep{chen2016from, lun2016s, lun2017nocounts, lund2012detecting}.
Similar to limma, we set the design matrix to be {\tt
  model.matrix({\textasciitilde}0 + cdr + conditions)}, where {\tt
  cdr} was the estimated cellular detection rate. Note that {\tt
  edgeR} does not provide estimates of the standard error, and
therefore cannot be used in combination with mash; see for example
\url{https://support.bioconductor.org/p/61640}. We found
that edgeR typically took much longer to run than limma or MAST.

\subsubsection{mash}
\label{sec:mash}

We used the mashr package (version 0.2.59;
\citealt{urbut2019flexible}) to fit a multivariate adaptive shrinkage
(``mash'') model to condition-level expression estimates produced by
limma. Specifically, we applied a variant of mash, ``common baseline
mash'', which is intended for testing and estimating differences
relative to a common reference point such as a control condition
\citep{sarah-thesis, yuxin-thesis}.

We took the following steps to fit the common baseline mash model and
recover mash posterior quantities:
\setlength\itemsep{0.4em}
\begin{enumerate}
  
\item {\em Normalize and transform counts.} 
We computed log-normalized counts $\tilde{y}_{ji}$ as
\begin{equation*}
\tilde{y}_{ji} = \log(0.1 + y_{ji} \times \med\{\tilde{s}_1, \ldots,
\tilde{s}_n \}/ \tilde{s}_i),
\end{equation*}
in which $\tilde{s}_i \coloneqq \sum_{j=1}^J y_{ji}$ is the size
factor for cell $i$. This was guided by the discussion in Chapter 2 of
\cite{jason-thesis}.

\item {\em Run limma.} We ran limma on the log-normalized counts
  $\tilde{y}_{ji}$ as before (Section~\ref{sec:limma}),
  except that here we set {\tt trend = TRUE}, {\tt robust = FALSE}.

\item {\em Define mash inputs.} The inputs to mash were the $J
  \times R$ matrix of condition-level expression estimates for all
  genes and the corresponding $J \times R$ matrix of standard
  errors. Since the conditions were randomly shuffled in the
  simulations, there was no correlations in the expression
  measurements among conditions, so we set $\bm C$ to be the $R \times
  R$ identity matrix (which is the default setting in {\tt
    mash\_set\_data}).

\item {\em Define prior covariance matrices.} mash requires
  the covariance matrices ${\bm U}_k$ in the prior to be
  specified beforehand. We defined a total of $K = R + 11$ prior
  covariance matrices, in which some were ``canonical'' covariance
  matrices, and others were ``data-driven'' matrices
  \citep{urbut2019flexible}.
\begin{enumerate}
\vspace*{0.4em}

\item {\em Define canonical covariance matrices.} First, we defined
  the $R + 5$ canonical covariance matrices ${\bm U_k}$ by calling
  mashr function {\tt cov\_canonical} with its default settings. This
  collection of covariance matrices included covariances capturing
  independent effects across all conditions (identity matrix); equal
  effects across all conditions (a matrix of ones); and
  condition-specific effects.

\item {\em Estimate data-driven covariance matrices.} Additionally, we
  estimated 6 data-driven covariance matrices using the Extreme
  Deconvolution (ED) method \citep{ed}. ED is included with mashr and
  called via the mashr function {\tt cov\_ed}. The ED estimates were
  initialized from a PCA of the effect estimates: a rank-5 covariance
  matrix defined by the top 5 PCs; and 5 rank-1 matrices each defined
  by one of the top 5 PCs. This was accomplished by calling mashr
  function {\tt cov\_pca} with {\tt npc = 5}. Both initial covariance
  estimates (via {\tt cov\_pca}) and final covariance estimates (via
  {\tt cov\_ed}) were obtained using only the strongest signals; that
  is, genes $j$ with $\mbox{\em lfsr} < 0.05$ in at least one
  condition as determined by a simple condition-by-condition analysis
  implemented in the ashr R package \citep{stephens2017false}.  Note
  that the ED updates are ``subspace preserving''; when the ED
  estimates are initialized as rank-1 matrices, the final estimates
  are also rank-1.

\item {\em Modify data-driven covariance matrices.} To deal with the
  issue that the {\em lfsr} can be underestimated when low-rank prior
  covariance matrices are used, we added a small scalar, $0.01 \times
  a_k$, to the diagonal of each data-driven covariance matrix ${\bm
    U}_k$, in which $a_k$ was the maximum value along the diagonal of
  ${\bm U}_k$.

\item {\em Determine scaling coefficients.} The mash prior also
  requires specification of the scaling coefficients $w_l$, and for
  this we used the automated selection of scaling coefficients
  implemented in mashr.
  
\end{enumerate}
  
\item {\em Fit mash model.} We fit the mash model and computed
  posterior quantities by calling the {\tt mash} function from mashr,
  in which all optional arguments were kept at their defaults. The one
  exception is that we set {\tt alpha = 1} in {\tt mash\_set\_data},
  which specifies the ``exchangeable {\em z}-scores'' (EZ) model
  \citep{stephens2017false, urbut2019flexible}.
  In initial trials, we found that the EZ model performed better than
  the ``exchangeable effects'' (EE) model.

\item {\em Detect expression differences.} We reported expression
  differences for gene-condition pairs based on the {\em lfsr}, and we
  identified differentially expressed genes based on the {\em minimum
    lfsr}.
  
\end{enumerate}

\subsubsection{Poisson mash}
\label{sec:poisson_mash_details}

The Poisson mash analysis pipeline was modeled after the mash
analysis pipeline.
We took the following steps to fit a Poisson mash model (or Poisson
mash RUV model) and compute key posterior quantities. Some steps
were specific to the Poisson mash RUV model. All steps were
implemented in the poisson.mash.alpha R package (version 0.1-87).
\begin{enumerate}
  
\item {\em Compute size factors.} We computed the cell-specific size
  factors $\tilde{s}_i$ from ${\bm Y}$ by calling function {\tt
    calculateSumFactors} from the scran R package, in which clusters
  were determined by running the {\tt quickCluster} function on ${\bm
    Y}$.
  
\item {\em Aggregate single-cell data.} We summed the UMI counts
  ${\bm Y}$ across cells by condition to obtain condition-level
  counts ${\bm X}$. This step was implemented in function {\tt
    pois\_mash\_set\_data}.

\item {\em Estimate unwanted variation from single-cell data (Poisson
  mash RUV only).} We estimated the $J \times D$ matrix ${\bm F}$ by
  fitting a GLM-PCA model to ${\bm Y}$ using the algorithms
  implemented in the {\tt glmpca} package (version 0.2.0.9000;
  \citealt{townes2019feature, glmpca}). Specifically, we fit a GLM-PCA
  model with negative-binomial likelihood ({\tt fam = "nb2"}) and with
  a separate overdispersion parameter for each gene.  We included the
  $R$ conditions as covariates in the GLM-PCA model. We used $D = 4$
  factors for the smaller simulated data sets and $D = 5$ factors for
  the larger data sets.

\item {\em Get initial parameter estimates.} To get sensible initial
  estimates of the parameters ${\bm\mu}$ and ${\bm\psi}$ (and
  ${\bm\rho}$ for Poisson mash RUV), we fit the Poisson mash model
  with the assumption that there were no differences in expression
  among the conditions; that is, $\beta_{jr} = 0$ for all $j = 1,
  \ldots, J$, $r = 1, \ldots, R$. Without expression differences, the
  mash prior is not needed, which simplifies estimation. These initial
  parameter estimates were then used to initialize subsequent model
  fitting. This step was implemented by function {\tt
    pois\_mash\_ruv\_prefit}.
  
\item {\em Determine covariance matrices in mash prior.} Like mash, we
  defined the covariance matrices ${\bm U}_k$ in the mixture prior in
  a separate step before fitting the full Poisson mash model. As in
  mash, We included a mixture of canonical and data-driven matrices in
  the prior. In total, we included $K = R + 7$ covariance matrices.
  
\begin{enumerate}
\vspace*{0.4em}
    
  \item {\em Define canonical covariance matrices.} For the canonical
    matrices, we included only the matrices capturing
    condition-specific effects. We did not include the ``independent
    effects'' matrix (the identity matrix) because it is not needed
    when the random effect is included. We did not include the ``equal
    effects'' matrix (a matrix of ones) since the $\mu_j$'s already
    perform a similar role.

  \item {\em Get initial estimates of data-driven covariance
    matrices.} As in the mash analysis pipeline, the data-driven
    covariance matrices were estimated using a two-step process: an
    initialization phase followed by a refinement phase. The
    initialization was performed by function {\tt pois\_cov\_init},
    and is analogous to the {\tt cov\_pca} step in the mash analysis
    pipeline (and indeed {\tt pois\_cov\_init} uses PCA in a very
    similar way to {\tt cov\_pca}). Instead of computing PCs using the
    limma effect estimates, we computed PCs using the {\em z}-scores
    from a multinomial goodness-of-fit test.  Specifically, for a
    given gene $j$, we tested whether $\lambda_{jr} = \lambda_j$ for
    all conditions $r = 1, \ldots, R$.  As in the mash analysis
    pipeline, we only used the genes $j$ containing ``strong
    signals.'' We selected strong signals based on the {\em z}-scores
    from the multinomial goodness-of-fit test; specifically, genes
    with at least one {\em z}-score greater than 3 in magnitude.
        
  \item {\em Refine estimates of data-driven covariance matrices.} The
    refinement phase was implemented in function {\tt pois\_cov\_ed}
    and is analogous to the {\tt cov\_ed} function in mash (noting
    that the function {\tt pois\_cov\_ed} is confusingly named and
    does not actually use Extreme Deconvolution). To refine the
    data-driven covariance matrices, we fit a simplified Poisson mash
    (or Poisson mash RUV) model without the scaling factors (i.e., $L
    = 1$, $w_1 = 1$). As in the initialization phase, only the genes
    with the strongest signals were used.
    
  \item {\em Modify data-driven covariance matrices.} Similar to the
    mash pipeline, after rescaling each ${\bm U}_k$ so that the
    largest entry along the diagonal was 1, we added a small
    constant, $\epsilon = 0.01$, to the diagonal entries of the
    data-driven covariance matrices. This was done to avoid possible
    issues with {\em lfsr} calculations.

  \item {\em Determine scaling factors in mash prior.} The mash prior
    also includes scaling factors $w_l$ to allow for a wide range of
    effect sizes. We used the automated setting of the scaling factors
    implemented in the {\tt pois\_mash} function. In brief, {\tt
      pois\_mash} chooses an evenly spaced grid of scaling factors, in
    which the range is determined dynamically by the data. A
    maximum-likelihood estimate of $\lambda_{jr}$ was computed for
    each gene $j$ and condition $r$ by $\hat{\lambda}_{jr} = (x_{jr} +
    0.1)/s_r$ (a pseudocount of 0.1 was added to avoid logarithms of
    zero). Then the range of the scaling factors was chosen based on
    the range of $\log \hat{\lambda}_{jr}$. We set {\tt gridmult =
      2.5} so that the scaling factors were spaced (multiplicatively)
    by a factor of 2.5.
    
\end{enumerate}

\item {\em Fit Poisson mash model.} We called {\tt pois\_mash} to fit
  the Poisson mash (or Poisson mash RUV) model. This function fits the
  model by iterating the variational EM updates. We ran at most 300
  model fitting iterations.
  
\item {\em Detect expression differences.} Finally, we reported a
  gene-condition pair $j, r$ as having an expression difference if the
  {\em lfsr} outputted by {\tt pois\_mash} was less than a specified
  threshold. We reported differentially expressed genes as those in
  which the {\em minimum lfsr} was less than a specified threshold.

\end{enumerate}

\subsection{Bulk RNA-seq data simulations with multiple replicates}
\label{sec:bulk-sims}

We took the following steps to simulate bulk RNA-seq data sets. We
generated RNA-seq count data for $4 \times R$ samples measured in
10,000 genes using the function {\tt generateSyntheticData} from the R
package compcodeR \citep{soneson2014compcoder}. This function
simulates the count for gene $j$ in sample $i$ from a negative
binomial distribution with mean parameter $s_i
\frac{\mu_j}{\sum_{j=1}^m \mu_j}$ and dispersion parameter $\phi_j$,
independently across samples and genes. The sequencing depth $s_i$ for
sample $i$ was drawn uniformly from the interval $[0.7 \times 10^7,
  1.4 \times 10^7]$, and $\mu_j, \phi_j$ were randomly drawn from
paired values estimated from real data
\citep{pickrell2010understanding, cheung2010polymorphic}.
Next, we randomly assigned each of the $4 \times R$ samples to one of
the $R$ conditions so that there were 4 replicate samples per
condition and no systematic expression differences among the
conditions. Finally, we used binomial thinning \citep{gerard2020data}
to simulate expression differences in 1,000 randomly chosen genes as
was described in Section \ref{sec:design} of the main text.

For each setting of $R$, we followed this procedure to simulate 20
data sets.

\subsection{Computing environment}

All analyses of simulated and real data sets were performed in R 4.1.0
\citep{R}, linked to the OpenBLAS 0.3.13 optimized numerical
libraries, on Linux machines (Scientific Linux 8) with Intel Xeon
E5-2680v4 (``Broadwell'') processors. Note that fitting the Poisson
mash models to the larger cytokines data sets required more memory, as
much as 100 GB, and for these larger data sets we also used
multithreading to speed up the computations (at most 3 cores).

\section{Cytokines experiment protocols and data preparation}
\label{sec:protocols}

\subsection{Mice samples}
\label{sec:samples}

Female C57BL/6J mice (Strain \#000664) were obtained from Jackson
Laboratories. Age-matched female mice (6 to 8 weeks old) were used in
all experiments. Animals were housed in specific pathogen-free and
BSL2 conditions at the University of Chicago. All experiments were
performed in accordance with the US National Institutes of Health
Guide for the Care and Use of Laboratory Animals and approved by the
University of Chicago Institutional Animal Care and Use Committee.

\subsection{Library preparation}

Mice were intravenously injected with \SI{2.5}{\micro\gram} of the
following 48 recombinant cytokines: IL-1$\beta$ (211-11B), IL-2
(212-12), IL-3 (213-13), IL-4 (214-14), IL-5 (215-15), IL-6 (216-16),
IL-7 (217-17), IL-9 (219-19), IL-10 (210-10), IL-11 (220-11), IL-12
p70 (210-12), IL-13 (210-13), IL-15 (210-15), IL-17A (210-17), IL-17F
(210-17F), IL-21 (210-21), IL-22 (210-22), IL-33 (210-33),
IFN-$\gamma$ (315-05), M-CSF (315-02), GM-CSF (315-03), G-CSF
(250-05), TNF (315-01A), TGF-$\beta$1 (100-21), CCL2 (250-10), CCL3
(250-09), CCL4 (250-32), CCL11 (250-01), CCL17 (300-30), CCL20
(250-27), CCL22 (250-23), CXCL1 (250-11), CXCL5 (300-22), CXCL9
(250-18), CXCL10 (250-16), CXCL12 (250-20A), CXCL13 (250-24)
(purchased from Peprotech), IFN-$\alpha$1 (751802), IFN-$\beta$
(581302) (purchased from BioLegend), IL-1$\alpha$ (400-ML-025), IL-18
(9139-IL-010), IL-23 (1887-ML-010), IL-25 (7909-IL-010), IL-27
(2799-ML-010), IL-34 (5195-ML-010), IL-36$\alpha$ (7059-ML-010), CCL5
(478-MR-025), and TSLP (555-TS-010) (purchased from R\&D
Systems). Untreated mice were used as a control. Three mice were used
for each cytokine treatment group and control group. Mice were
anesthetized with avertin (250-500 \si[per-mode =
  symbol]{\milli\gram\per\kilo\gram}) \SI{12}{\hour} after cytokine
injection and whole blood was collected from the left ventricle of the
heart into the tubes containing ice-cold buffer (PBS plus 0.5\% FCS
and \SI[per-mode = symbol]{2}{\milli\mole\per\liter} EDTA). Red blood
cells were lysed in ACK Red blood cell lysis buffer (Lonza, BW10548E)
for \SI{3}{\minute} and then 3 biological replicates per treatment
group were pooled into a tube. After pooling samples, the cells were
washed by the buffer. Anti-Ter119 MicroBeads (Miltenyi Biotec,
130-049-901) was used to deplete Ter119$^+$ cells thoroughly. Cells
were then label<ed with the TotalSeq\textsuperscript{TM}-C0301
anti-mouse Hashtag 1 antibody (BioLegend, 155861) and
TotalSeq\textsuperscript{TM}-C0302 anti-mouse Hashtag 2 antibody
(BioLegend, 155863) at \SI{0.25}{\micro\gram}/sample for
\SI{15}{\minute} at \SI{4}{\degreeCelsius}. After staining, cells were
washed with PBS, and resuspended in PBS + 0.04\% BSA. Cells were
counted and resuspended at a density of 1000
cells/\si{\micro\liter}.

Cell capture and library preparations were performed using the 10x
Genomics Chromium controller, the Chromium Single Cell 5’ Library \&
Gel Bead Kit (10x Genomics, PN-1000006), the Chromium Single Cell 5’
Feature Barcode Library Kit (10x Genomics, PN-1000080), the Chromium
Chip A Single Cell Kit (10x Genomics, PN-120236), the Chromium i7
Multiplex Kit (10x Genomics, PN-120262), and the Chromium i7 Multiplex
Kit N, Set A (10x Genomics, PN-1000084) according to the
manufacturer’s protocol. Libraries were loaded onto a NextSeq 500/550
High Output Kit v2.5 (75 Cycles) (Illumina, 20024906). Hashtag
indexing was used for demultiplexing the sequencing data. Sequencing
was done in two batches, with 3 cytokine treatments (TNF,
IFN-$\alpha$1, IFN-$\beta$) plus one control in the first batch and 45
cytokine treatments plus one control in the second batch.

In our analyses, we used only the data from the
second batch to avoid batch effects. We also excluded the IL-27
due to a technical issue in the experiment.

\subsection{Raw read processing}
\label{sec:preprocessing}

Raw sequencing data were processed by the Cell Ranger pipeline
\citep{zheng2017massively} to produce UMI counts. Quality control
steps were taken to filter out cells with too low ($<$400) or high
($>$22,000) counts, as well as cells with a large fraction of
mitochondrial counts ($>$20\%).

Cells were assigned to cell types by clustering using the Louvain
algorithm \citep{blondel2008fast}. Based on known marker genes, we
annotated 8 of the clusters as cell types: B cells ({\em Cd19}, {\em
  Cd79a}, {\em Cd79b}); CD4$^{+}$ T cells ({\em Cd3d}, {\em Cd3e},
{\em Thy1}, {\em Cd4}), CD8$^{+}$ T cells ({\em Cd3d}, {\em Cd3e},
{\em Thy1}, {\em Cd8a}); dendritic cells ({\em Flt3}); Ly6C$^{-}$
monocytes ({\em Nr4a1}, {\em Pparg}); Ly6C$^{+}$ monocytes ({\em
  Ly6c2}, {\em F13a1}), neutrophils ({\em S100a8}, {\em S100a9}); and
natural killer (NK) cells ({\em Ncr1}). For a given cell type, genes
expressed in fewer than 20 cells were removed.

The processed data used in the Poisson mash analyses is
summarized in Supplementary Table \ref{table:cytokines-data-summary}
and in the Supplementary Data.



\clearpage

\noindent {\bf Supplementary figures.}

\begin{figure}[th]
\centering
\includegraphics[width=\textwidth]{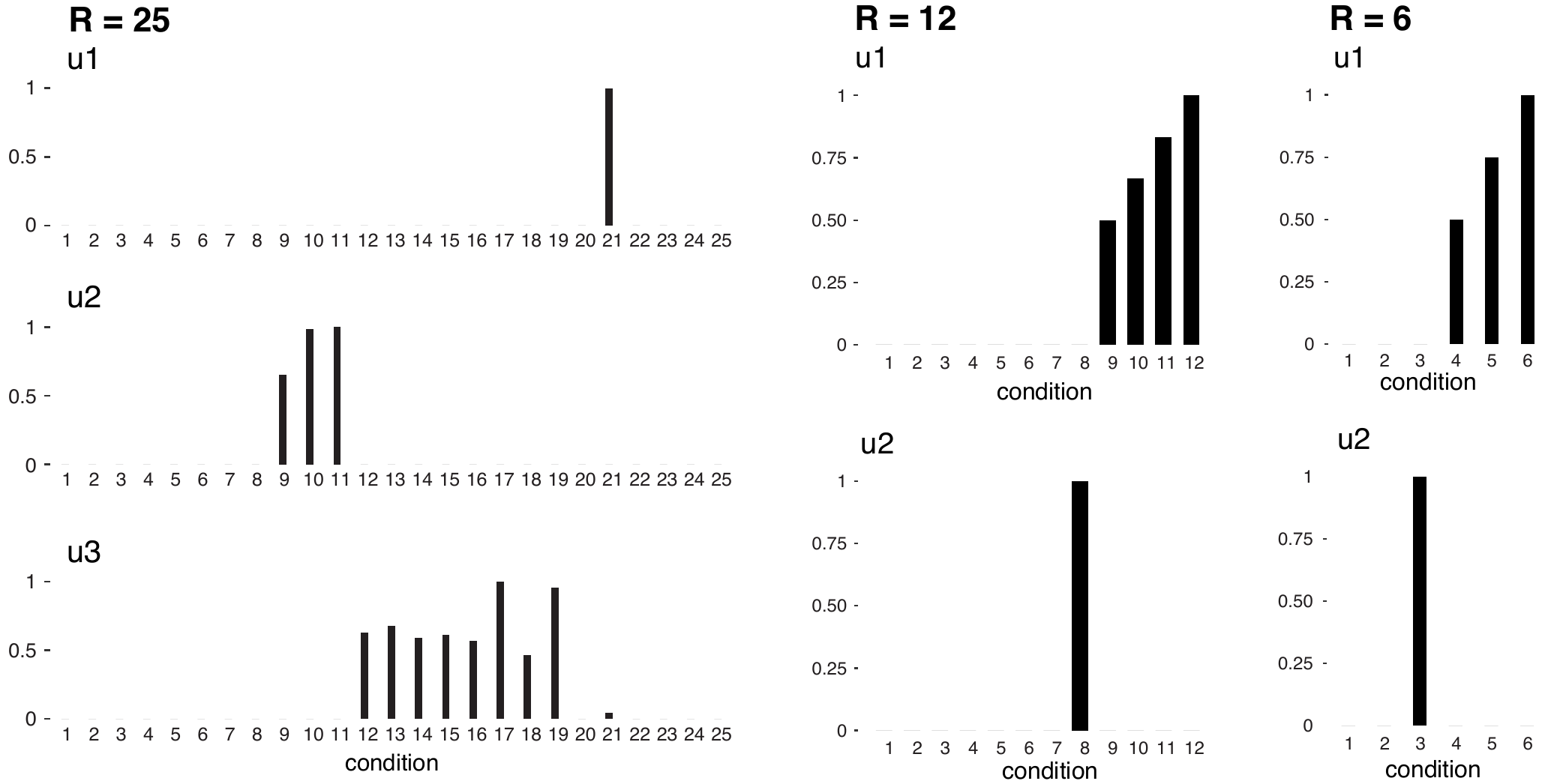}
\caption{\rm Cross-condition sharing patterns used to simulate the
  treatment effects for data sets simulated with $R = 25$, 12 and 6.}
\label{fig:design}
\end{figure}

\begin{figure}[th]
\centering \includegraphics[width=\textwidth]{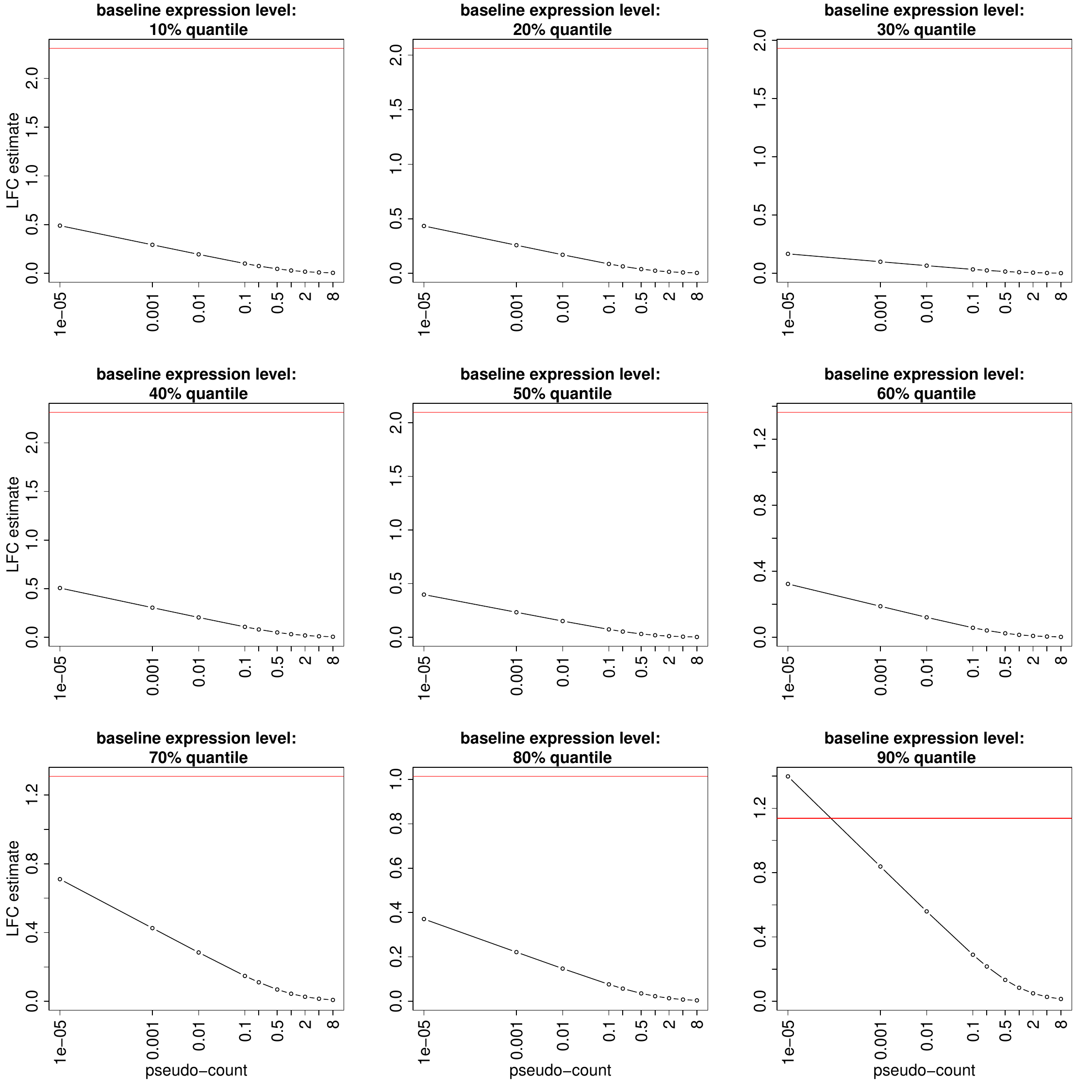}
\caption{\rm Illustration of impact of pseudocount on LFC estimation
  bias. Results show the LFC estimate obtained using limma (circles,
  joined by black line) against the truth (red horizontal line) for
  different choices of pseudocount. The 9 panels show results for 9 DE
  gene-condition pairs chosen from a simulated data set in the larger
  sample size scenario, with the genes ordered by their baseline
  expression level. These results are consistent with
  \cite{lun2018overcoming} who showed that differences in
  log-transformed expression values (with pseudocount) between
  conditions often do not accurately estimate the LFC, especially when
  counts are low.}
\label{fig:figure4}
\end{figure}

\begin{figure}[th]
\centering
\includegraphics[width=\textwidth]{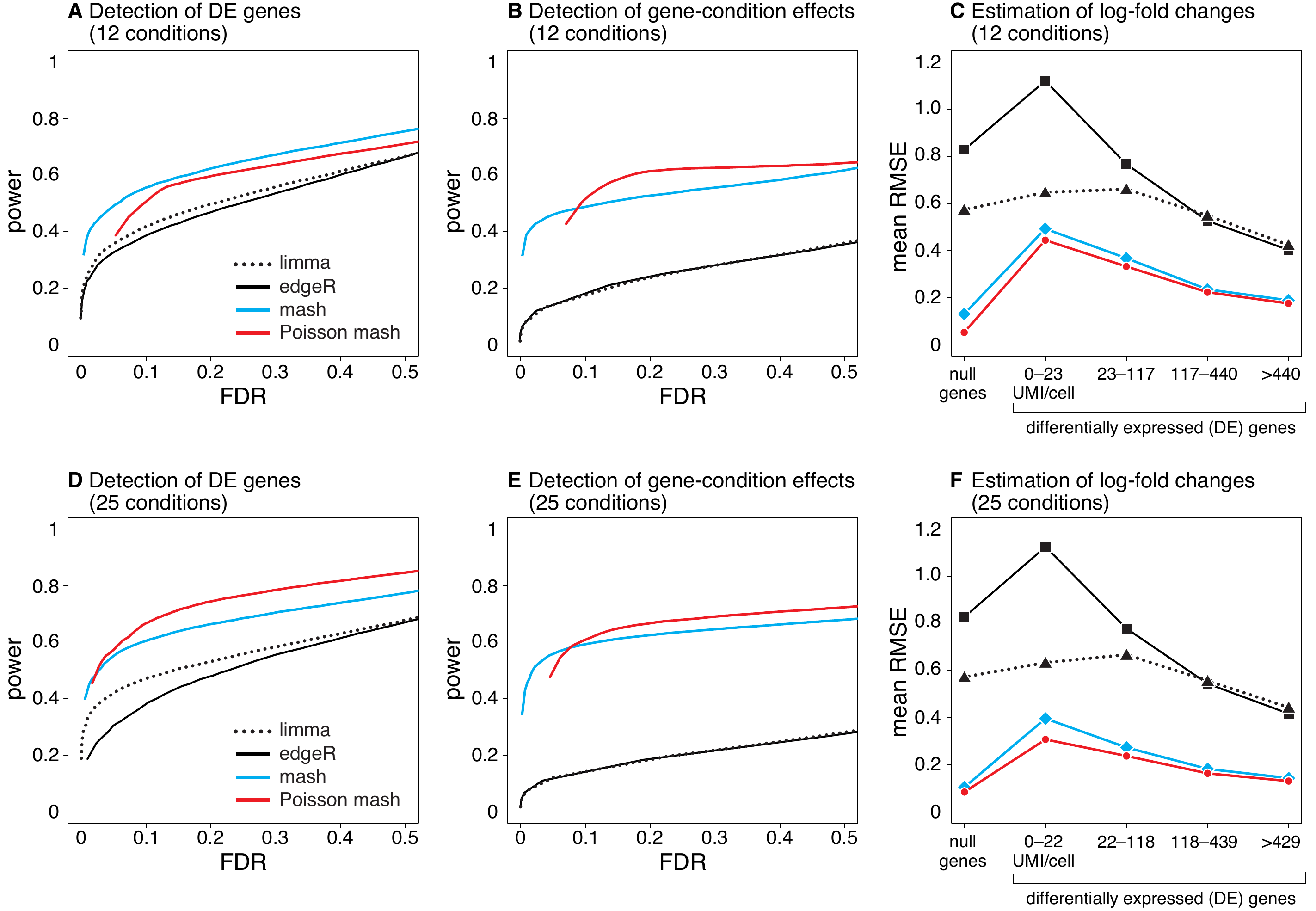}
\caption{\rm Evaluation of DE analysis methods in simulated
    bulk RNA-seq data sets, $R = 12$ conditions (top row) and $R = 25$
    conditions (bottom row). FDR and power were calculated for all
    genes (Panels A, D) and for all gene-condition pairs (B, E) in the
    20 simulations by varying a {\em p}-value or {\em lfsr} threshold
    from 0 to 1. Panels C and F summarize LFC estimation accuracy by
    the RMSE, averaged over 20 simulations. RMSE was calculated in
    non-overlapping groups of genes: ``null'' genes (genes in
    which there were no differences in expression in all conditions);
    and DE genes grouped by expression level (read counts per sample).}
\label{fig:sims_bulk}
\end{figure}

\begin{figure}[th]
\centering
\includegraphics[width=\textwidth]{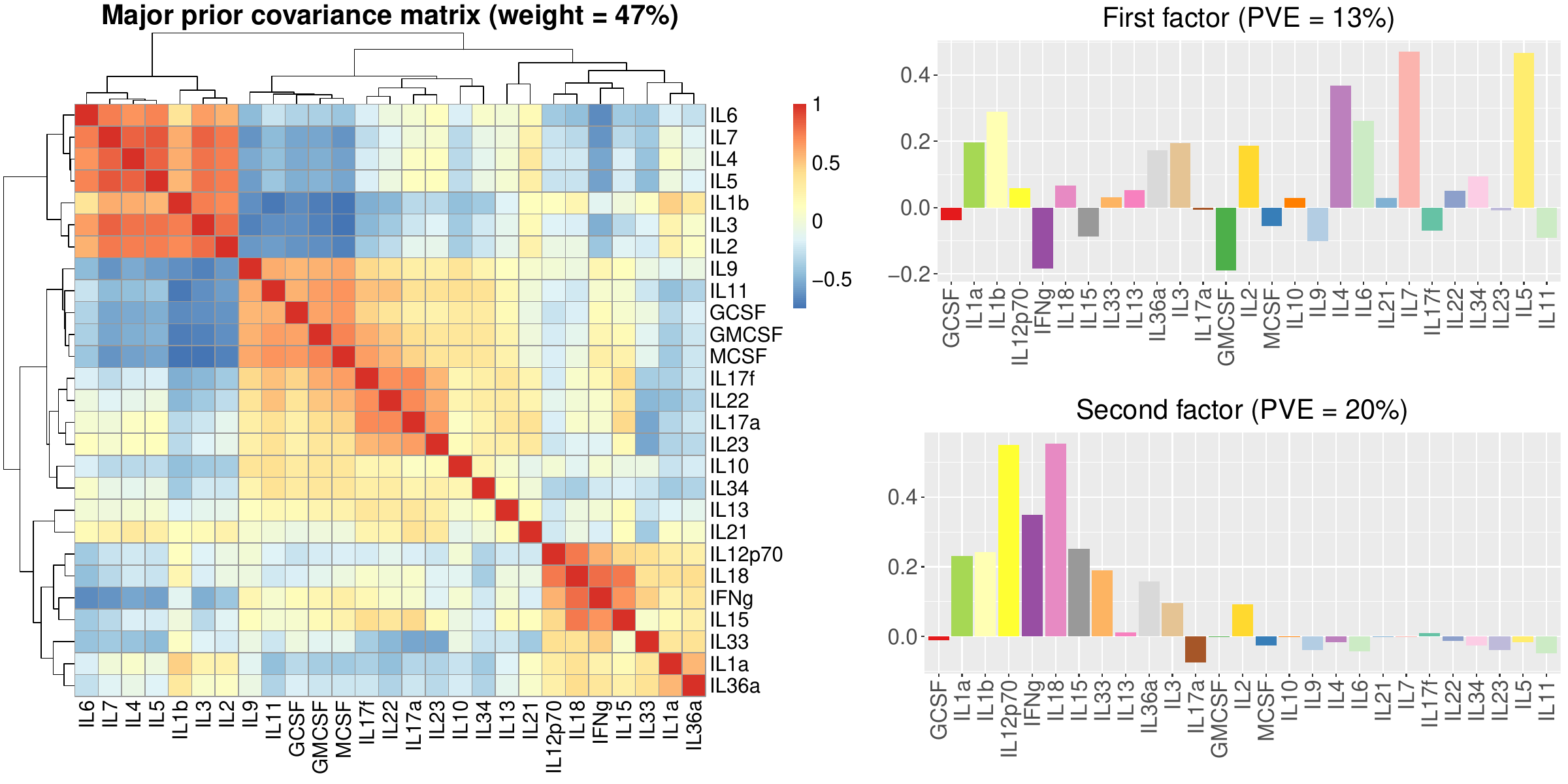}
\caption{\rm Poisson mash RUV estimates of prior (left) and posterior
  (right) patterns of differential expression across cytokine
  treatments in B cells. Left panel shows a heatmap of the correlation
  matrix of the prior covariance matrix that receives the largest
  weight in the Poisson mash RUV fit.  Right panel shows the top
  factors by PVE from a factor analysis of the posterior mean LFC
  estimates.}
\label{fig:bcells}
\end{figure}

\begin{figure}[th]
\centering
\includegraphics[width=\textwidth]{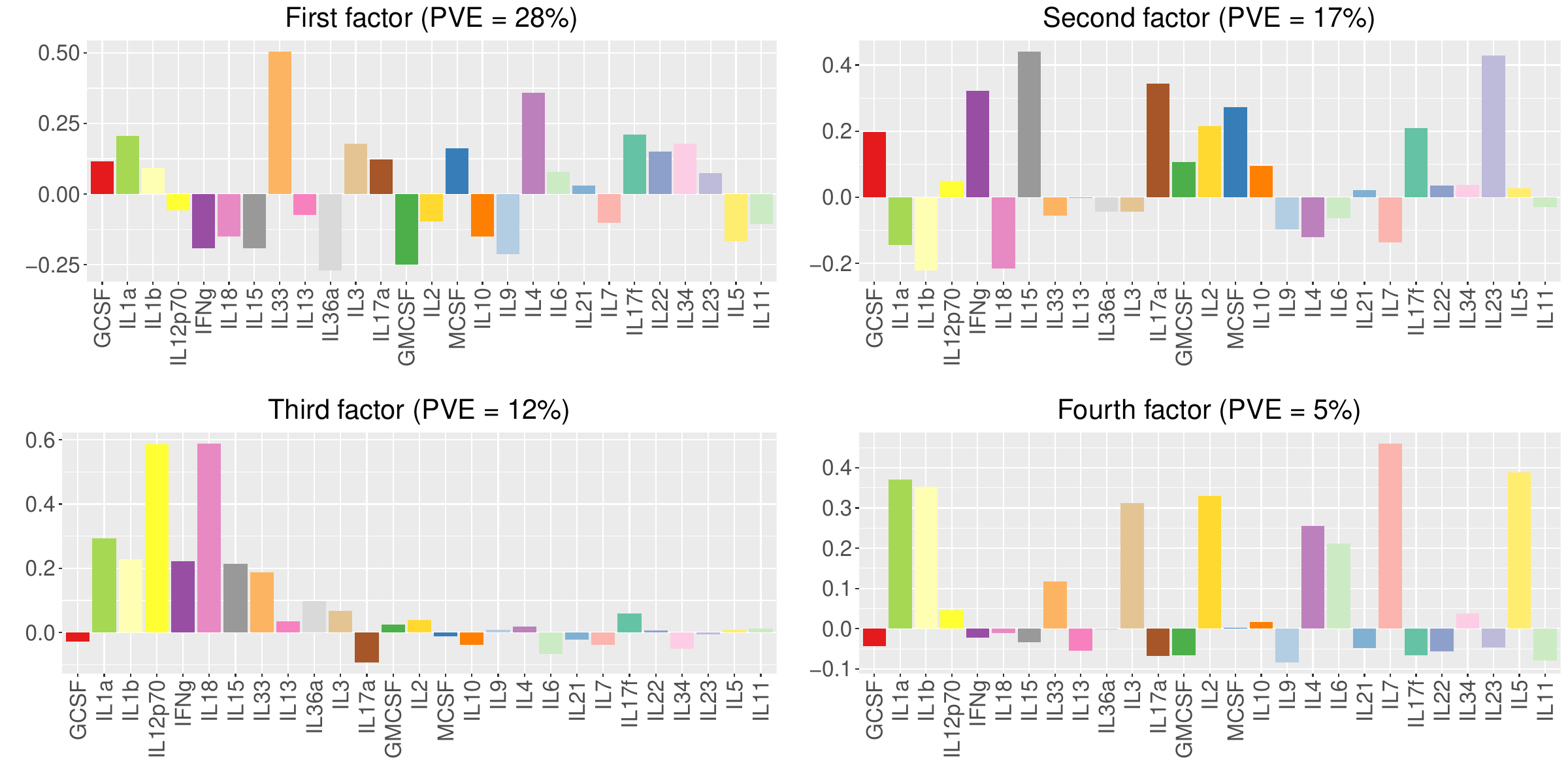}
\caption{\rm Poisson mash RUV estimates of posterior patterns of
  differential expression across cytokine treatments in CD4$^{+}$ T
  cells. The top factors by PVE from a factor analysis of the
  posterior mean LFC estimates are shown.}
\label{fig:cd4}
\end{figure}

\begin{figure}[th]
\centering
\includegraphics[width=\textwidth]{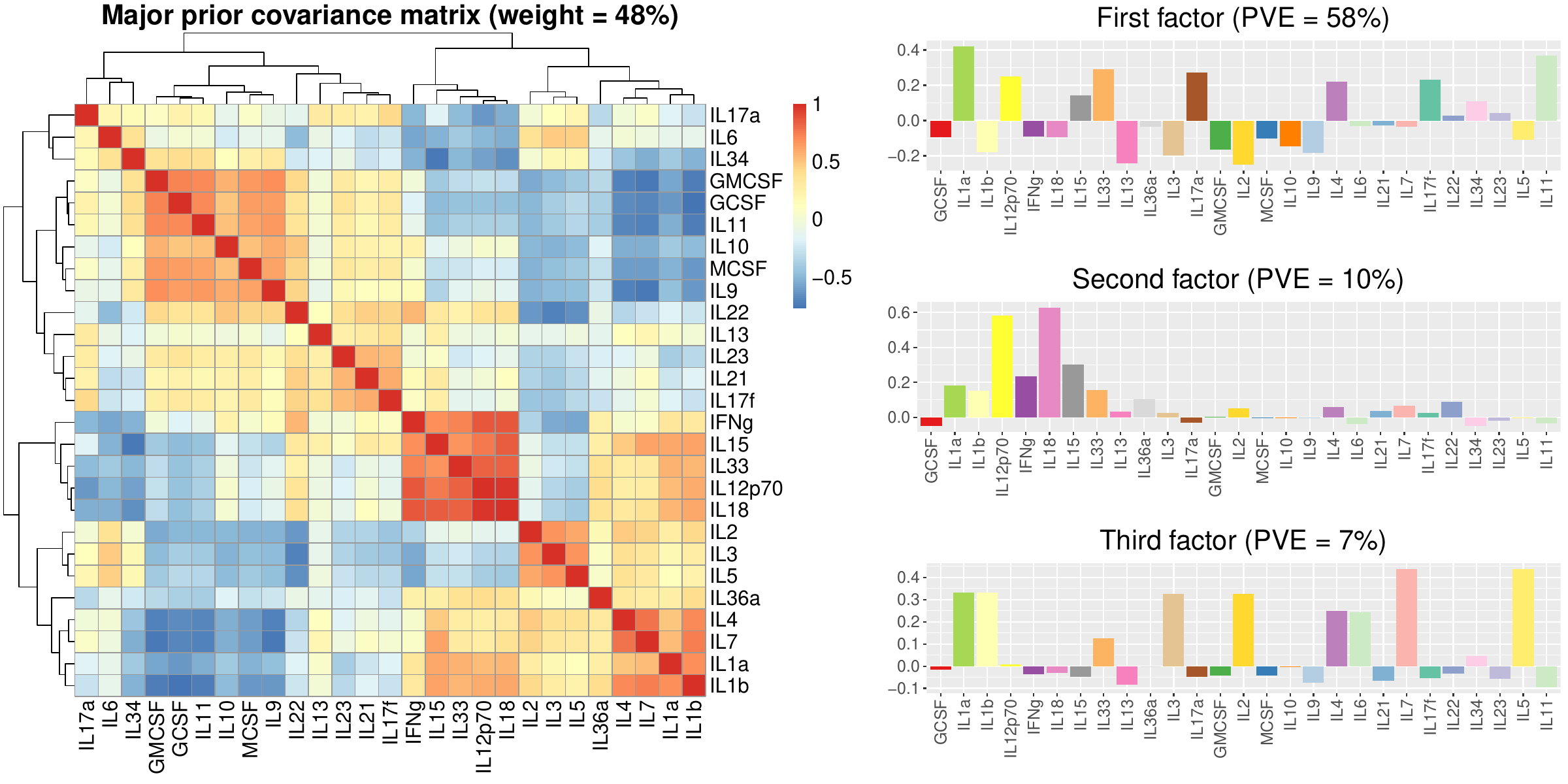}
\caption{\rm Poisson mash RUV estimates of prior (left) and posterior
  (right) patterns of differential expression across cytokine
  treatments in CD8$^{+}$ T cells. Left panel shows a heatmap of the
  correlation matrix of the prior covariance matrix that receives the
  largest weight in the Poisson mash RUV fit.  Right panel shows the
  top factors by PVE from a factor analysis of the posterior mean LFC
  estimates.}
\label{fig:cd8}
\end{figure}

\begin{figure}[th]
\centering
\includegraphics[width=\textwidth]{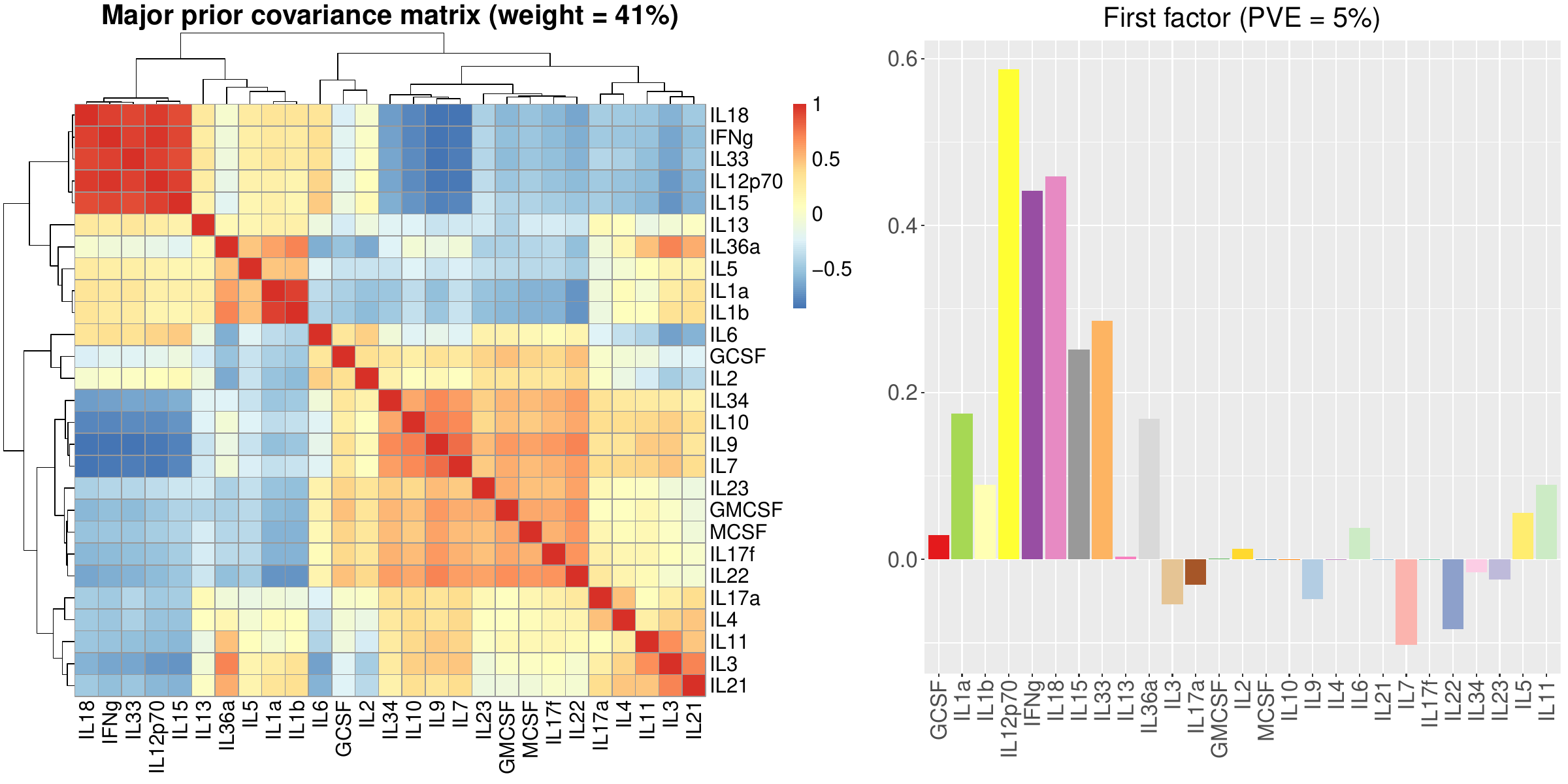}
\caption{\rm Poisson mash RUV estimates of prior (left) and posterior
  (right) patterns of differential expression across cytokine
  treatments in dendritic cells. Left panel shows a heatmap of the
  correlation matrix of the prior covariance matrix that receives the
  largest weight in the Poisson mash RUV fit.  Right panel shows the
  top factor by PVE from a factor analysis of the posterior mean LFC
  estimates.}
\label{fig:dendritic}
\end{figure}

\begin{figure}[th]
\centering
\includegraphics[width=\textwidth]{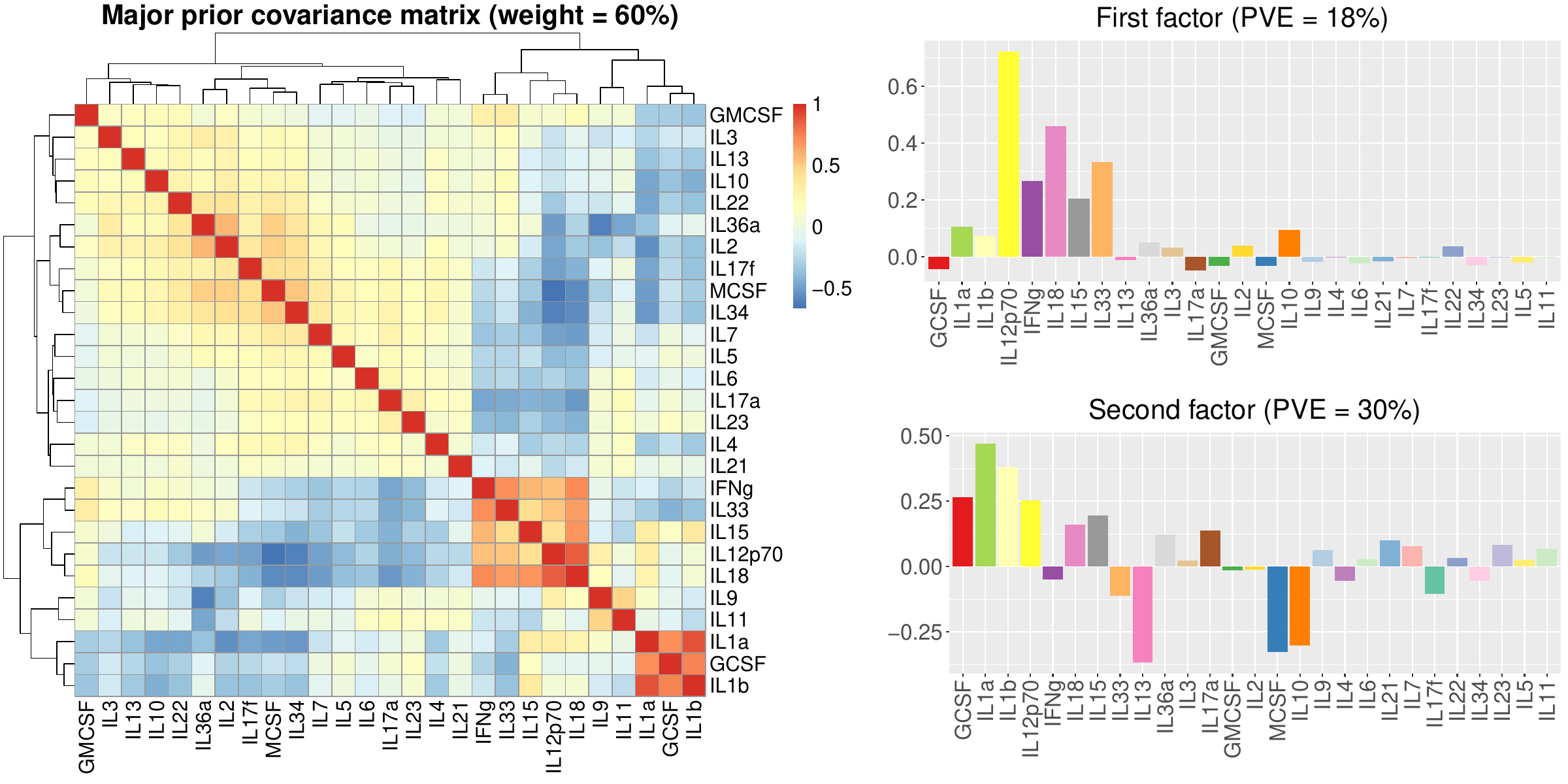}
\caption{\rm Poisson mash RUV estimates of prior (left) and posterior
  (right) patterns of differential expression across cytokine
  treatments in Ly6C$^{-}$ monocytes. Left panel shows a heatmap of the
  correlation matrix of the prior covariance matrix that receives the
  largest weight in the Poisson mash RUV fit.  Right panel shows the
  top factors by PVE from a factor analysis of the posterior mean LFC
  estimates.}
\label{fig:ly6c-}
\end{figure}

\begin{figure}[th]
\centering
\includegraphics[width=\textwidth]{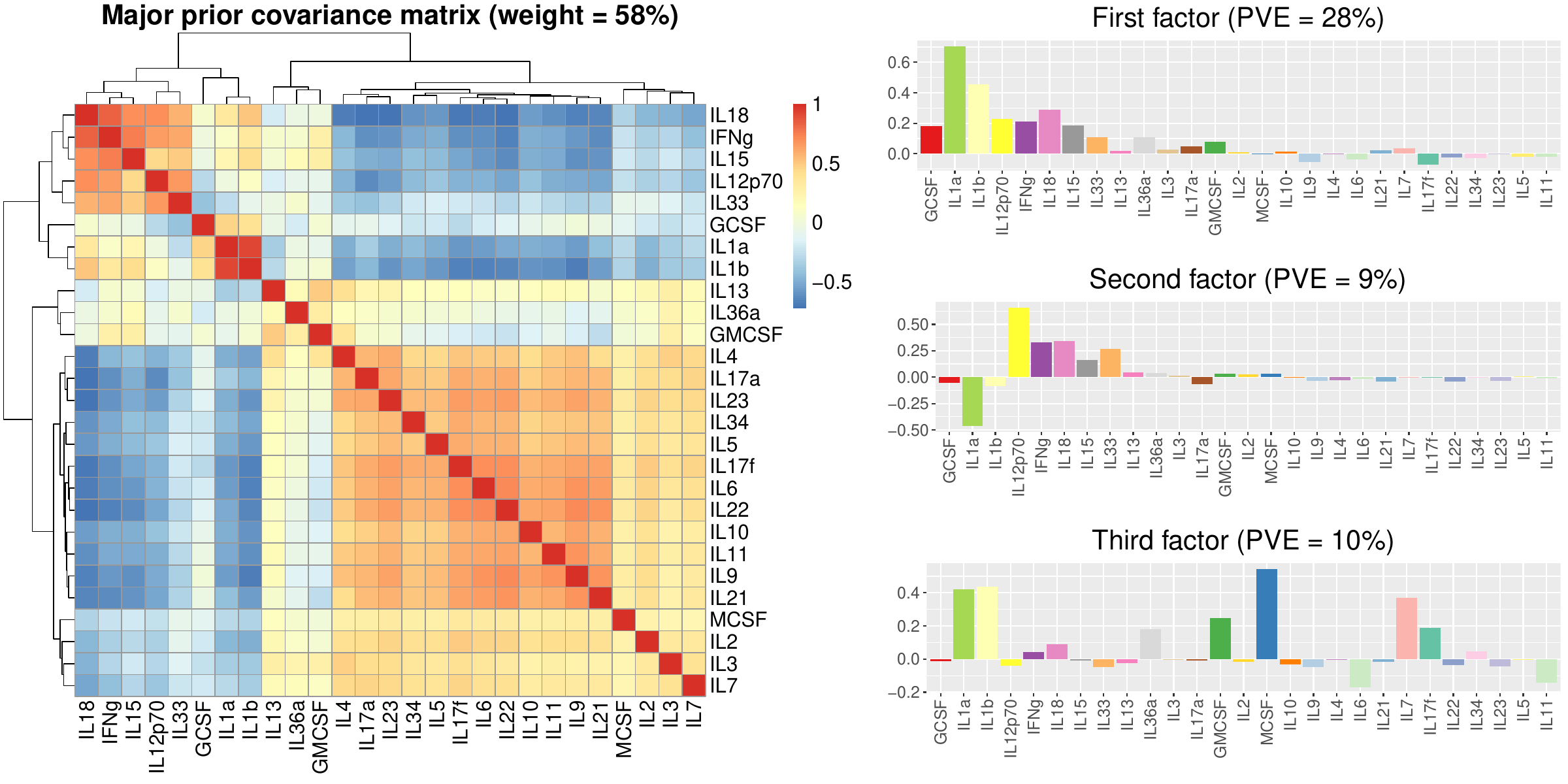}
\caption{\rm Poisson mash RUV estimates of prior (left) and posterior
  (right) patterns of differential expression across cytokine
  treatments in Ly6C$^{+}$ monocytes. Left panel shows a heatmap of the
  correlation matrix of the prior covariance matrix that receives the
  largest weight in the Poisson mash RUV fit.  Right panel shows the
  top factors by PVE from a factor analysis of the posterior mean LFC
  estimates.}
\label{fig:ly6c+}
\end{figure}

\begin{figure}[th]
\centering
\includegraphics[width=\textwidth]{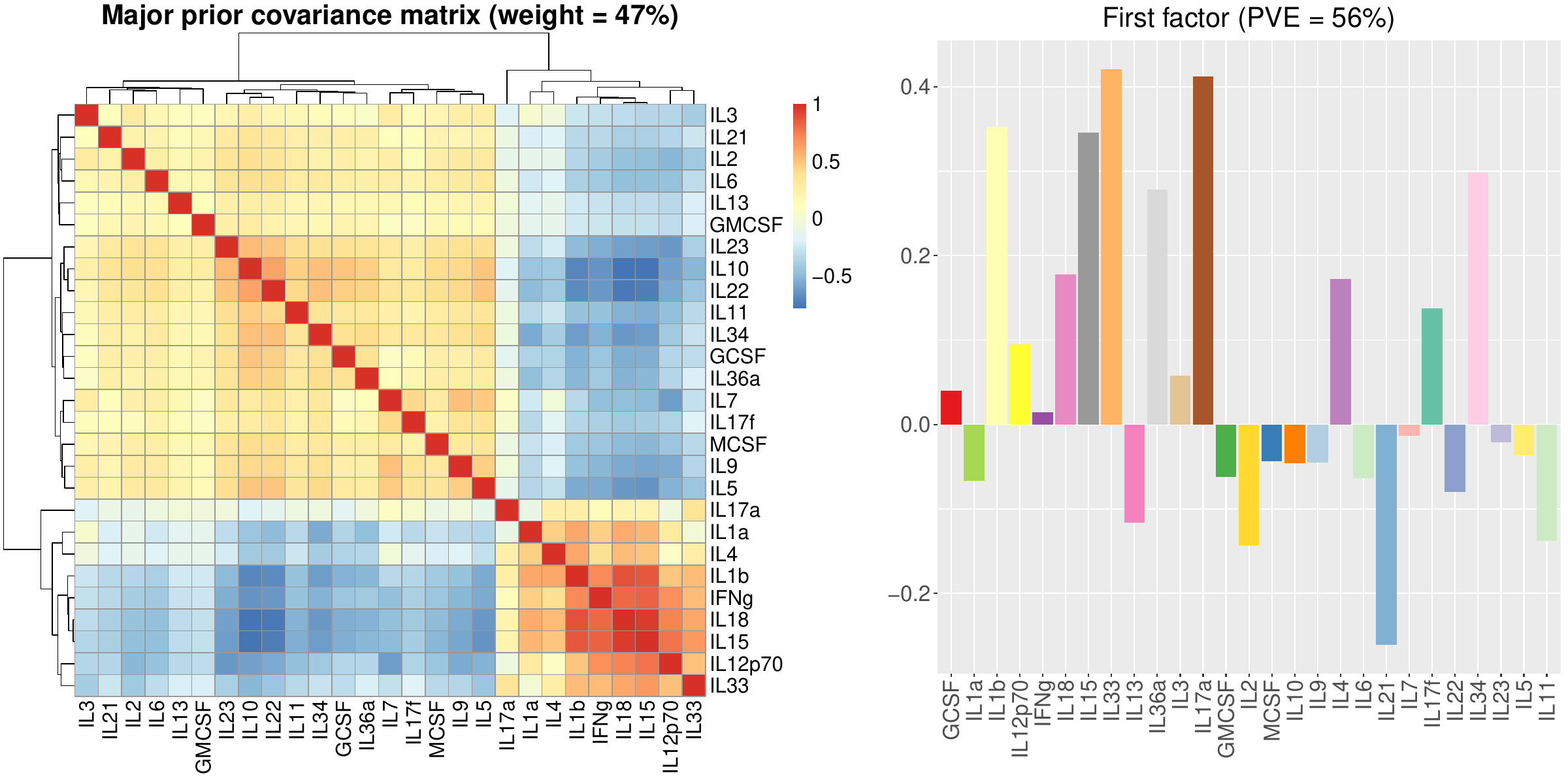}
\caption{\rm Poisson mash RUV estimates of prior (left) and posterior
  (right) patterns of differential expression across cytokine
  treatments in NK cells. Left panel shows a heatmap of the
  correlation matrix of the prior covariance matrix that receives the
  largest weight in the Poisson mash RUV fit.  Right panel shows the
  top factor by PVE from a factor analysis of the posterior mean LFC
  estimates.}
\label{fig:nk}
\end{figure}

\begin{figure}[th]
\centering
\includegraphics[width=\textwidth]{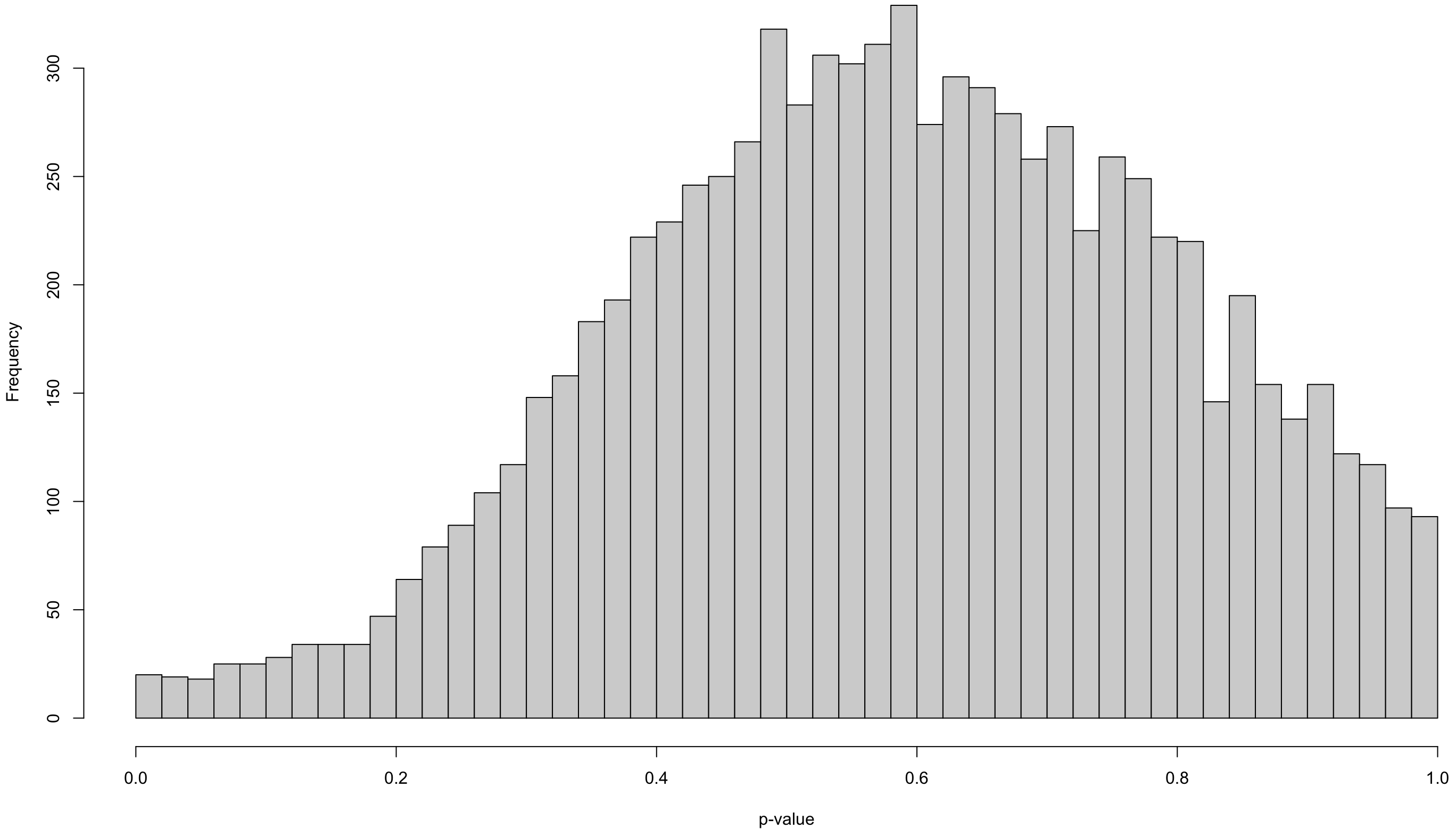}
\caption{\rm Assessment of goodness-of-fit of the Poisson mash RUV model
  fit to the neutrophils data. The histogram shows the empirical
  distribution of {\em p-}values from the goodness-of-fit tests for
  8,543 genes.}
\label{fig:gof}
\end{figure}

\clearpage

\noindent {\bf Supplementary tables.}

\begin{table}[th]
\caption{\rm Summary of the processed cytokine stimulation data. ``Median
  total UMI count'' is the median of the cell-wise total UMI counts.}
\label{table:cytokines-data-summary}
\setlength{\tabcolsep}{36pt}
\begin{tabular*}{\linewidth}{ @{} *{5}c @{}}
\toprule
& & median total \\
cell type & cells &
UMI count & genes \\
\midrule
B cells & 87,140 & 1,789 & 11,698 \\
CD4$^{+}$ T cells & 12,939 & 1,971 & 10,462 \\
CD8$^{+}$ T cells & 14,282 & 2,791 & 10,819 \\
dendritic cells & 778 & 6,255 & 8,952 \\  
Ly6C$^{-}$ monocytes & 3,677 & 3,824 & 10,236 \\
Ly6C$^{+}$ monocytes & 4,229 & 5,269 & 10,404 \\
neutrophils & 13,362 & 1,127 & 8,543 \\
natural killer cells & 1,735 & 2,194 & 8,496 \\
\bottomrule
\end{tabular*}
\end{table}

\begin{table}[th]
{\small
  \caption{\rm Patterns of differential expression shared across cell
    types based on Gene Ontology (GO) gene set enrichment analyses. To
    generate the GO enrichment results using WebGestalt
    \citep{liao2019webgestalt, go, go-2020}, we first performed, for
    each cell type, a factor analysis \citep{wang2021empirical} on the
    $N_{\mathrm{de}} \times 27$ matrix of posterior mean LFCs from the
    $N_{\mathrm{de}}$ genes that were identified as differentially
    expressed by Poisson mash RUV in that cell type. We defined the
    ``driving genes'' for a factor as all genes with $\text{\em lfsr}
    < 0.001$, then we split the driving genes into two sets based on
    whether the factor estimate was positive or negative. We performed
    two GO gene set enrichment analyses for each factor: one for the
    positive driving genes, and one for the negative driving
    genes. The table lists the results for all ``significantly
    enriched'' GO gene sets which we defined as gene sets with a false
    discovery rate (FDR) less than 0.05. Columns in the table from
    left-to-right are: the cytokine conditions with the largest factor
    values; the cell types in which the factors were identified;
    whether the enrichment was based on the positive or negative
    driving genes; and the significantly enriched GO gene sets.}
\label{table:go-shared}
\begin{center}
\vspace*{1ex}
\begin{tabular}{cc@{}c@{}c@{}c}
\toprule
top cytokines & cell types & \makecell{direction of \\ regulation} &
significant GO terms \\
\midrule
\makecell{IL12p70, IL18, IFNg, \\ IL15, IL33} & \makecell{B cells,
  \\ CD4 T cells, \\ CD8 T cells, \\ Ly6C- monocytes, \\ Ly6C+
  monocytes, \\ neutrophils} & + &
\makecell{response to interferon-beta, \\ response to
  interferon-gamma, \\ response to virus, \\ antigen processing and
  presentation, \\ regulation of innate immune response, \\ regulation
  of immune effector process, \\ defense response to bacterium,
  \\ positive regulation of defense response, \\ positive regulation
  of cytokine production, \\ regulation of response to cytokine
  stimulus} \\
\midrule

\makecell{IL33, IL4 \\ \\ \\ IL12p70, IL17a, IL17f, \\ IL1a, IL4,
  IL11, \\ IL15, IL33 \\ \\ \\ IL17a, IL15, IL1b, \\ IL33, IL34,
  IL36a} &
\makecell{CD4 T cells, \\ \\ \\ \\ CD8 T cells
  \\ \\ \\ \\ NK cells \\ NK cells}
& + &
\makecell{chromosome localization, \\ chromosome segregation,
  \\ microtubule cytoskeleton organization \\ involved in mitosis,
  \\ spindle organization, \\ organelle fission, \\ regulation of
  chromosome organization, \\ DNA conformation change, \\ regulation
  of mitotic cell cycle, \\ positive regulation of cell cycle, \\ cell
  cycle phase transition} \\

\midrule
\makecell{IL1a, IL1b, IL2, IL3, \\ IL4, IL5, IL6, IL7 \\ \\ \\
  IL12p70, IL15, IL18, \\ IFNg, IL33} &
\makecell{B cells, \\ CD4 T cells, \\ CD8 T cells \\ \\
  dendritic cells, \\ Ly6C+ monocytes} &
\makecell{+ \\ \\ \\ - } &
\makecell{cytoplasmic translation, \\ ribonucleoprotein complex
  \\ subunit organization, \\ ncRNA metabolic process,
  \\ ribonucleoprotein complex biogenesis} \\
\bottomrule
\end{tabular}
\end{center}
}
\end{table}

\begin{table}[th]
{\small
  \caption{\rm Cell-type-specific patterns of differential expression
    based on Gene Ontology (GO) gene set enrichment analyses. See
    Supplementary Table \ref{table:go-shared} for details.}
\label{table:go-cell-type-specific}
\begin{center}
\vspace*{1ex}
\begin{tabular}{cc@{}c@{}c}
\toprule
top cytokines & \makecell{cell type}
& \makecell{direction of \\ regulation} & significant GO terms \\
\midrule
\makecell{IL1a, IL1b, \\ IL2, IL3, IL4, IL5, IL6, IL7} & B cells & - & \makecell{digestive system development, \\ peptidyl-threonine modification, \\ embryonic organ development, \\ cell-cell signaling by wnt, protein acylation, \\ peptidyl-serine modification, \\ cell surface receptor signaling pathway \\ involved in cell-cell signaling, \\ covalent chromatin modification, \\ peptidyl-lysine modification} \\
\midrule
\makecell{IL15, IFNg, IL23, \\ IL17a, IL17f, GCSF,  MCSF} & CD4 T cells & + & \makecell{coagulation, homotypic cell-cell adhesion, \\ regulation of tube size, \\ regulation of body fluid levels, \\ response to wounding, \\ fatty acid metabolic process} \\
\midrule
IL1a, IL1b, GCSF, IL12p70 & Ly6C- monocytes  & + & \makecell{humoral immune response, \\ response to interleukin-1, \\ negative regulation of proteolysis, \\ regulation of vasculature development, \\ cell chemotaxis, \\ regulation of peptidase activity} \\
\midrule 
\makecell{IL1a, IL1b, GCSF, \\ IL12p70, IL15, IL18, IFNg} & Ly6C+ monocytes & + & \makecell{mitochondrial gene expression, \\ protein folding, \\ protein localization to mitochondrion, \\ nucleoside monophosphate metabolic process, \\ nucleoside triphosphate metabolic process, \\ generation of precursor metabolites and energy, \\ purine-containing compound metabolic process, \\ ribose phosphate metabolic process} \\
\midrule
IL1a, IL1b, GCSF & neutrophils  & + & \makecell{morphogenesis of a polarized epithelium, \\ regulation of synapse structure or activity, \\ cell junction organization,  \\ transition metal ion homeostasis,  \\ synapse organization, \\ positive regulation of cytoskeleton organization, \\ actin filament organization, \\ regulation of supramolecular fiber organization, \\ regulation of actin filament-based process, \\ regulation of inflammatory response} \\
\bottomrule
\end{tabular}
\end{center}
}
\end{table}

\end{appendices}

\clearpage

\bibliographystyle{apalike}
\bibliography{references}

\end{document}